\documentclass[aps,prd,floatfix,nofootinbib,twocolumn,10pt]{revtex4}

\usepackage{ulem}
\usepackage[utf8]{inputenc}
\usepackage{amsmath}
\usepackage{amssymb}
\usepackage{graphicx,bm}
\usepackage[colorlinks,citecolor=blue]{hyperref}
\usepackage{slashed}

\usepackage{amsmath}
\usepackage{amsfonts}
\usepackage{natbib}
\usepackage{color}
\usepackage{epstopdf}
\usepackage{multirow}
\usepackage{longtable}
\usepackage{rotating}
\usepackage{url}
\usepackage{amssymb,mathrsfs}
\usepackage[indent=10pt]{parskip}

\usepackage{subfigure}
\usepackage{parskip}

\newcommand{\gag}{g_{a\gamma}}
\usepackage{amsmath}

\begin{document}

\title{
X-ray polarimetric features of Gamma-ray Bursts across varied redshifts and hints for Axion-Like-Particles
}

\author{Qingxiang Zhang$^{1}$, Feng Huang$^{1, }$\footnote{Corresponding author: fenghuang@xmu.edu.cn}, Zhongxiang Wang$^{2}$, Taotao Fang$^{1}$}

\affiliation{$^{1}$Department of Astronomy, Xiamen University, Xiamen, Fujian 361005, China \\
$^{2}$Department of Astronomy, School of Physics and Astronomy, Yunnan University, Kunming 650091, China}

\date{\today}

\begin{abstract}
Polarimetric features during the prompt phase of Gamma-ray Bursts (GRBs) have been essential for elucidating the debated emission mechanisms and gaining insight into the inner structure of GRBs. However, the potential impact of photon-Axion-Like-Particle (ALP) mixing in extragalactic magnetic fields, leading to significant modifications to the initial polarization state, has been overlooked in discussions concerning prompt phase constraints. In this work, we first examine the statistical characteristics of linear polarization degree ($\Pi_{L}$) in GRBs, by utilizing data from polarimetric missions focusing on sub-MeV emissions. Our analysis, conducted with a restricted sample of GRBs spanning various redshifts, reveals a diverse distribution of $\Pi_{L}$, which currently shows no correlation with the GRBs' spectral parameters or properties of candidate host galaxies. We then explore alternations to the initial $\Pi_{L}$ due to photon-ALP mixing within a domain-like structure of the intergalactic magnetic field (${\bf B}_{\rm IGM} $), considering various parameter sets associated with ALPs and ${\bf B}_{\rm IGM}$. With the existence of ALPs with mass $m_{a}$$~$$\lesssim$$~$$10^{-14}$$~$eV and photon-ALP coupling constant $g_{a\gamma}~$$\simeq$$~0.5\times10^{-11}$$~$GeV$^{-1}$,  
we  find that fully linearly polarized photons ($\Pi_{L_{0}}=1$) may experience a polarization reduction of up to $20\%$, whereas for unpolarized photons ($\Pi_{L_{0}}=0$), the mixing can increase polarization by up to ~$40\%$ for GRBs with redshifts above approximately 1. To ensure that the effect of mixing is small enough to be negligible, the mixing term $\Delta_{a\gamma} \equiv 1/2\  g_{a\gamma} {\bf B}_{\rm IGM}$ should be less than $1.5\times 10^{-4}$ Mpc$^{-1}$. Currently, the number of GRBs with both sub-MeV polarization measurement and redshift confirmation remains very limited. Half of the GRBs with available polarization data have $\Pi_{L}$ being less than 30\%. Certification of redshift for this subset of GRBs would further constrain the parameter space of low-mass ALPs or provide an independent means to determine the upper limit on ${\bf B}_{\rm IGM}$.

\end{abstract}


\maketitle

\section{Introduction}
\label{sec:intro}

GRBs are among the most energetic events observed in the universe and their prompt phase is characterized by a sudden and intense release of gamma-rays with energies spanning from tens of keV to several MeV. Despite considerable advancements in observation and theory, elucidating the origin of this emission remains a complex challenge due to the diverse nature and intricate physics involved. Ongoing research aims to decipher spectral features and temporal evolution across various emission mechanisms and geometric structures \citep{GRB-review-ZB2018}. While spectral data alone prove insufficient, polarimetric measurements are crucial in clarifying the prompt-emission nature and inner structure of GRBs. However, it is important to recognize that the measured value of polarization is typically ascribed to photons emitted by GRBs, hardly considering possible propagation effects in extragalactic space. Such effects could appreciably affect both the flux and polarization state of high-energy photons originating from distant energetic sources, particularly in models incorporating the existence of ALPs, which are very light spin-zero particles predicted by numerous extensions of the Standard Model \citep{2010ReviewALPs}.

Recently, ALPs have been increasingly highlighted as promising candidates for dark matter. ALPs are regarded as generalized forms of axions, exhibiting less stringent correlations between its mass and coupling constant compared to axions. Observations of high-energy astrophysical sources have sparked significant interest in narrowing down the parameter space of ALPs and seeking evidence of their existence. The recent observability of very-high-energy emission (above $10\,\rm TeV$, \citep{LHAASO10TeV}) from GRB 221009A, known as the Brightest Of All Time (BOAT), has been considered a potential indication of ALP existence \citep{galanti_PRL}. This observation offers a possible explanation for the significant decrease in the optical depth of TeV photons, attributed to photon-ALP mixing in external magnetic fields, a key characteristic of ALPs as described by the Lagrangian term
\begin{equation}
 {\cal L}_{a\gamma}=\gag \, {\bf E}\cdot{\bf B}\,a~,   
\end{equation}
where $\gag$ is the coupling constant of the photon-ALP mixing, {\bf E} and {\bf B} refer to the electric and magnetic fields respectively. 

Similar to this scenario of photon-ALP mixing, previous investigations concerning anomalies in the transparency of the universe have primarily centered on analyzing the spectral characteristics of TeV photons emitted from cosmological sources \citep{2007Mirizzi-TeV,2009Hints, Abramowski_2013, 2014JCAP...09..003M, liang2021effect, blazar_TeV,2024TeVwiggles}. Furthermore, potential spectral deviations in GeV emissions from blazars have been attributed to photon-ALP mixing in various magnetic fields, including those within jets, intracluster environments, intergalactic space, and the Milky Way \citep{Tavecchio2012_FSRQ, Ajello_2016, Zhangcun2018, huang2021, NGC1275_2021, lihaijun2022, Meyer_flat-spectrum_radio_quasars}. Corresponding constraints on the photon-ALP coupling constant $\gag$ have been derived for the mass of ALPs ($m_a$) ranging from $10^{-10}$ eV to $10^{-7}$ eV. For a recent comprehensive review of these constraints, refer to Ref.$~$\citep{Galanti_2022review} and references therein. 

In addition to case studies examining the effects of photon-ALP mixing, larger samples at cosmological distances could offer a statistical perspective on this issue. The scatter in the luminosity relations of a substantial sample of active galactic nuclei (AGN) has been utilized to search for ALPs \citep{2009AGN-luminosity}. 
Ref.$~$\citep{2009AngelisAGNz-first} 
identified a correlation between spectral slope and redshift in the observed $\gamma$-ray spectra of blazars, assuming a consistent intrinsic emission spectrum for sources at different redshifts, and this phenomenon has been effectively explained in Ref.$~$\citep{2020Gal-blazar-z} through the scenario of photon-ALP mixing in intergalactic magnetic fields for blazars with redshifts up to 0.6. A more recent study has further investigated this correlation with a sample of blazars with redshift up to 1, confirming the necessity of photon-ALP mixing to explain the correlation within the redshift range of $0.2$ to $1$ \citep{2023Dong-AGN-z}.

Due to the spin-zero nature of ALPs, their mixing with photons in external magnetic fields not only affects the flux of photons but also alters their polarization states. The early spike in interest examining the effects of this mixing on the polarization state of photons stemmed from the reasonable explanation 
for observations of polarization in quasars at optical wavelengths. The observed large-scale alignment of quasar polarizations at different redshifts \citep{2001quasarpol-first,2002quasar-galaxy-Alignments,2011Nishant-quasar-Alignments,2012Payez-quasar-pol} and optical circular polarization observed in several quasars have been considered as hints for the existence of ALPs \citep{2012Payez-quasar-pol}. More recently, based on the absence of detectable optical circular polarization from blazars at a level of 0.1$\%$, Ref.$~$\citep{2023PhRvD.107d3031Y} have placed constraints on the ALPs-photon mixing term, limiting it to $g_{a\gamma} \cdot \bf B_{\rm jet}$ $\lesssim$ $7.9\times10^{-12}$ G $\cdot$ GeV$^{-1}$ for axion masses $m_{a}$ $\lesssim$ $10^{-13}$ eV ($\bf B_{\rm jet}$ refers to the magnetic field in the jet of blazar). For even low-energy photons, the mixing of ALPs with cosmic microwave background (CMB) photons has tended to influence the intensity of the CMB, as well as its linear and circular polarization, leading to distortions of the CMB spectrum, which subsequently imposed strong constraints on ALP parameters with the latest observations from CMB \citep{2008SignatureCMB,2012CMB}. For higher-energy photons, theoretical exploration of the propagation of X-ray to $\gamma$ ray photons originating from the central region of a cluster of galaxies or within the jet of a blazar has been undertaken in a series of articles \citep{galanti_pol_initial, galanti_cluster_pol, 2023GalAGN-jet-pol, galanti_pol_review}. An expected feature related to our purpose is that unpolarized photons from the source at a cosmological distance become partially polarized due to the photon-ALP mixing, possibly resulting in measurable modifications to the initial $\Pi_{L}$.

The concept of utilizing polarized prompt emission from GRBs to constrain ALP parameters has been investigated in several studies. 
In Ref.$~$\citep{Sakharov-GRBpol}, photon-ALP mixing was hypothesized to occur in the inner jet region, with empirical investigations placing constraints on $g_{a\gamma}\lesssim 10^{-12}$ GeV$^{-1}$ for ALPs with $m_a=10^{-3}$ eV. Another study focused on photon mixing in the internal shocks region of GRBs with a magnetic field strength of $10^{6}$ G, revealing alterations in both the early-time evolution pattern of $\Pi_{L}$ and the spectral shape compared to synchrotron radiation models \citep{GRBpol-jet}. Additionally, a comprehensive examination of the modification of photon polarization emitted from GRBs due to extragalactic magnetic fields was conducted, demonstrating that the existence of ALPs could contribute to broadening the polarization distribution of photons \citep{GRBpol-IGM}. Similar conclusions were drawn in Ref.$~$\citep{GRBpol-IQUV}, where two different configurations of the intergalactic magnetic field (IGM) ${\bf B}_{\rm IGM}$ were considered. Their analytical and numerical calculations  indicated that $\Pi_{L}$ tends towards an asymptotic value as the distance increases, which remains independent of the assumed magnetic field configuration.

The above earlier research primarily focused on the impact of photon-ALP mixing from the theoretical phenomenological model discussions, given the lack of polarimetric detection techniques at higher energies. However, this situation has changed in the past decades, particularly within the sub-MeV energy range. Early-stage space-borne X-ray polarimeters, such as IPS (100 keV–1 MeV) and IBIS (15 keV–10 MeV) on the INTEGRAL satellite, have contributed to the advancement \citep{INTEGRAL_IBIS2003}. Moreover, the first two dedicated GRB polarimeters, namely the Gamma-Ray Burst Polarimeter (GAP, 70--300 keV) \citep{GAP2011} and POLAR (50--500 keV) \citep{2020A&A...644A.124K}, have recorded numerous detections of GRB polarization. The AstroSAT satellite is presently operational with a polarimeter (CZTI) covering energies from 100 keV to 600 keV \citep{Chattopadhyay2022}, and it has also released a relatively large catalog of GRBs with polarization measurements. In this article, we will 
start with analyzing the statistical characteristics of $\Pi_{L}$ released by these 
X-ray polarimetric missions, and investigate modifications to the initial $\Pi_{L}$ resulting primarily from photon-ALP mixing within a domain-like structure of the intergalactic magnetic field, following the main framework of Ref.$~$\citep{GRBpol-IGM}.

The paper is organized as follows. In Sec.$~$\ref{Sec_Statistical}, we analyze the statistical properties of $\Pi_{L}$ data obtained by X-ray polarimetric missions. In Sec.$~$\ref{Sec_mixing}, we delve into the fundamental framework of photon-ALP mixing in external magnetic fields. In Sec.$~$\ref{Sec_results}, we present results related to modifications of the initial linear polarization degree $\Pi_{L}$ observed in GRBs, considering a range of ALP and magnetic field parameters. Finally, Sec.$~$\ref{Sec_conclusion} encompasses the conclusions and discussions.

\section{Statistical characteristics of $\Pi_{L}$}
\label{Sec_Statistical}

\subsection{$\Pi_{L}$ Distribution }
\label{sample}

\begin{figure}[!tp]
\begin{minipage}[t]{0.99\linewidth}
\centering
\includegraphics[width=1.0\textwidth]{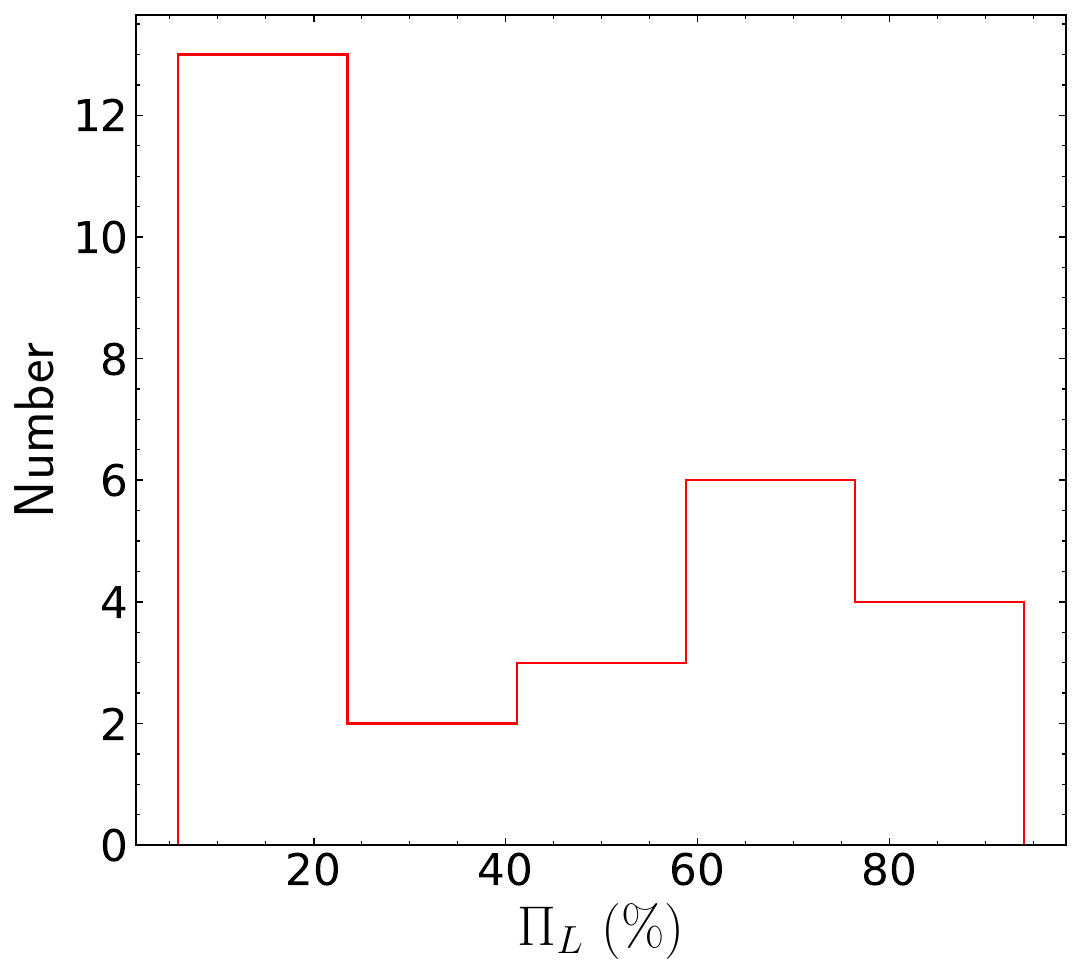}
\caption{Histogram of  $\Pi_{L}$  for 28 sources with the detected value.}
\label{fig:pol_hist}
\end{minipage}
\end{figure}

\begin{figure*}[htp]
\begin{minipage}[t]{0.99\linewidth}
\centering
\includegraphics[width=1.0\textwidth]{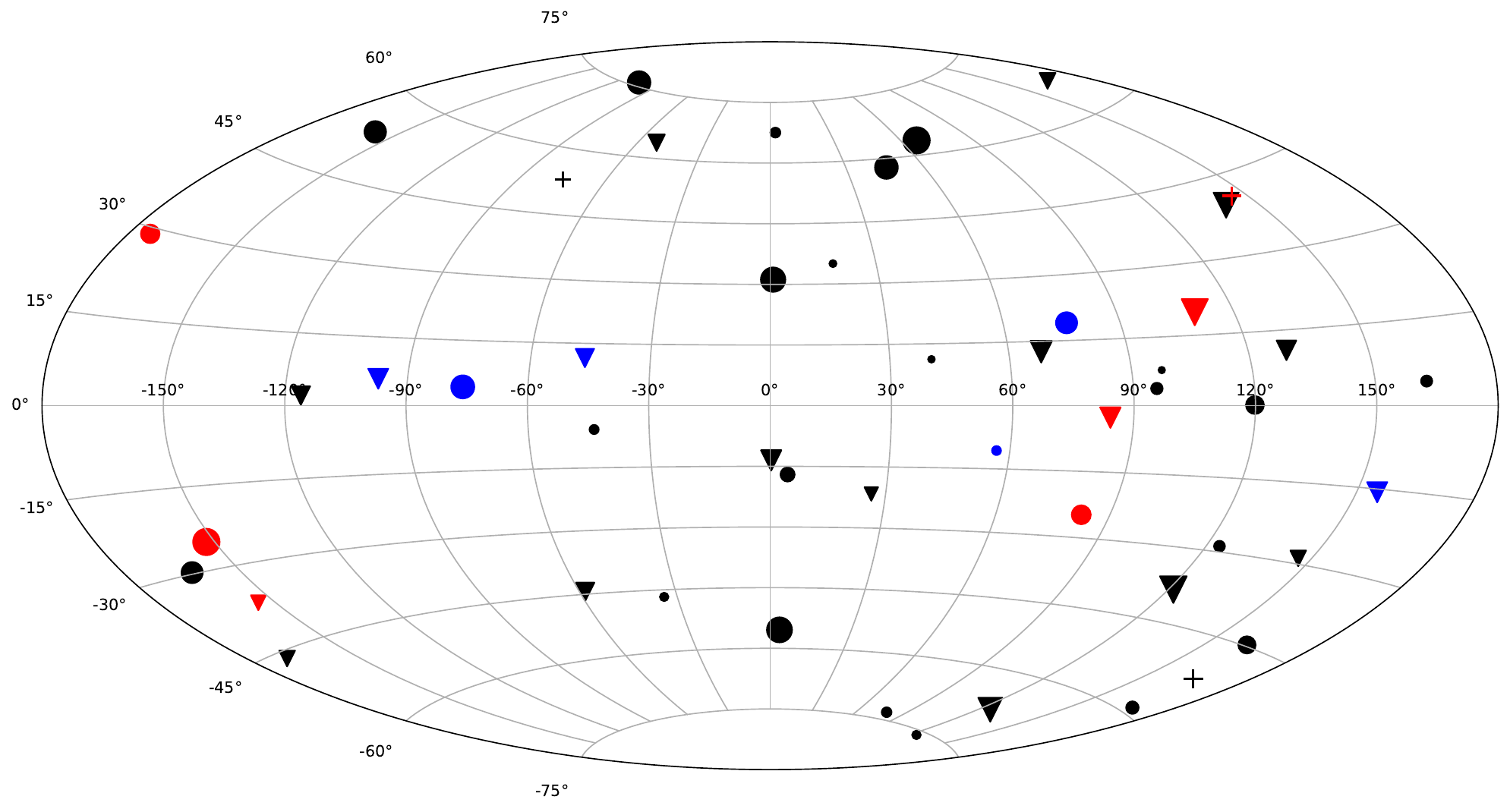}
\caption{All-sky distribution of 50 GRBs in Galactic coordinates. Dots indicate sources with measured $\Pi_{L}$. Inverted triangles represent sources with upper limits on $\Pi_{L}$, and plus signs represent lower limits. The red/blue data points represent sources with spectroscopic/photometric redshifts. Symbol size reflects the relative magnitude of $\Pi_{L}$.}

\label{fig:GRB_distribution}
\end{minipage}
\end{figure*}

\begin{table*}
\centering
        \resizebox{\textwidth}{!}{
            \begin{tabular}{c c c c c c c c c c c}
                \hline
                  &GRB & $\Pi_{L}~$($\%$) & $\alpha_{s}$ & $\beta_{s}$ & $E_{p}$ (keV)  & Fluence (10$^{-6}~$erg$~$cm$^{-2}$) &Instrument &Redshift$^{*}$  &Offset\\
                \hline\noalign{\smallskip}
                1 & 200311A  & $<45.41$& $-0.95_{-0.02}^{+0.02}$& $-2.57_{-0.19}^{+0.19}$ & $1218_{-110}^{+110}$ &$42.543_{-0.12789}^{+0.12789}$ & AstroSAT$^{\mathrm{I}}$ & $0.0838^{+0.0349}_{-0.0302}$$^{a}$ & $0.072^{\prime}$/8.00kpc\\[5pt]
                2 & 180103A  & $71.43_{-26.84}^{+26.84}$& $-1.31_{-0.06}^{+0.06}$& $-2.24_{-0.13}^{+0.90}$ & $273_{-23}^{+26}$ &223 & AstroSAT$^{\mathrm{I}}$ & $0.037^{+0.0063}_{-0.0015}$$^{a}$ & $0.184^{\prime}$/8.75kpc\\[5pt]
                3 & 180427A  & $60.01_{-22.32}^{+22.32}$& $-0.29_{-0.08}^{+0.08}$ & $-2.80_{-0.16}^{+0.16}$ & $147_{-2}^{+2}$  &$50.455_{-0.12559}^{+0.12559}$ & AstroSAT$^{\mathrm{I}}$ & $0.0309^{+0.045}_{-0.0309}$$^{a}$ & $0.273^{\prime}$/10.81kpc\\[5pt]
                4 & 200412A  & $<53.84$& $-0.70_{-0.05}^{+0.05}$& $-2.50_{-0.21}^{+0.21}$ & $256_{-7}^{+8}$ &$28.750_{-0.097405}^{+0.097405}$ & AstroSAT$^{\mathrm{I}}$ & $0.1055^{+0.0192}_{-0.0145}$$^{a}$ & $0.131^{\prime}$/18.74kpc\\[5pt]
                5 & 061122A  & $11_{-11}^{+48}$& $-1.14_{-0.32}^{+0.27}$ & $-1.91_{-0.10}^{+0.07}$ & $70_{-63}^{+106}$  &20 & INTEGRAL$^{\mathrm{II}}$  & $1.33_{-0.76}^{+0.77}$$^{b}$ & $0.435^{\prime \prime}$/20.13kpc\\[5pt]
                6 & 200806A  & $<54.73$& $-0.53$& $-2.96$ & 109.12 &1 & AstroSAT$^{\mathrm{I}}$ & $0.1148^{+0.1749}_{-0.1148}$$^{a}$ & $0.234^{\prime}$/36.58kpc\\[5pt]
                7 & 190530A  & $46.85_{-18.53}^{+18.53}$& $-0.99_{-0.00}^{+0.02}$& $-3.50_{-0.25}^{+0.25}$ & $888_{-8}^{+8}$ &$370.62_{-0.052475}^{+0.052475}$ & AstroSAT$^{\mathrm{I}}$  & $0.9386$$^{c}$ &-\\[5pt]
                8 & 180914B  & $48.48_{-19.69}^{+19.69}$& $-0.75_{-0.04}^{+0.04}$& $-2.10_{-0.70}^{+0.08}$ & $453_{-24}^{+26}$ &598 & AstroSAT$^{\mathrm{I}}$  & $1.096$$^{d}$ &-\\[5pt]
                9 & 171010A  & $<30.02$& $-1.12_{-0.00}^{+0.01}$& $-2.39_{-0.02}^{+0.02}$ & $180_{-3}^{+3}$ &$632.79_{-0.098525}^{+0.098525}$ &  AstroSAT$^{\mathrm{I}}$  & 0.3285$^{d}$ &-\\[5pt] 
                10 & 160703A  & $<62.64$& $-0.78_{-0.09}^{+0.12}$& $\textless -2.48$ & $351_{-46}^{+40}$ &9 & AstroSAT$^{\mathrm{I}}$  & $<1.5$$^{d}$ &-\\[5pt]
                11 & 160623A  & $<56.51$& $-0.94_{-0.02}^{+0.02}$& $-2.83_{-0.10}^{+0.09}$ & $662_{-18}^{+19}$ &$3.9564_{-0.068702}^{+0.068702}$ & AstroSAT$^{\mathrm{I}}$  & $0.367$$^{d}$ &-\\[5pt]
                12 & 160509A  & $<92$& $-0.75_{-0.02}^{+0.02}$ & $-2.13_{-0.03}^{+0.03}$ & $334_{-10}^{+12}$  &$178.98_{-0.14957}^{+0.14957}$ & AstroSAT$^{\mathrm{III}}$  & $1.17$$^{d}$ &-\\[5pt]
                13 & 160131A  & $94_{-33}^{+33}$& $-1.16_{-0.04}^{+0.04}$ & $-1.56_{-0.10}^{+0.07}$ & $586_{-259}^{+518}$ &20.4 & AstroSAT$^{\mathrm{III}}$  & $0.972$$^{d}$ &-\\[5pt]
                14 & 140206A  & $>48$& $-0.94_{-0.08}^{+0.08}$ & $-2.0_{-0.30}^{+0.20}$ & $98_{-17}^{+17}$  &$15.520_{-0.074778}^{+0.074778}$ & INTEGRAL$^{\mathrm{IV}}$  & $2.739_{-0.001}^{+0.001}$$^{e}$ &-\\[5pt]
                \hline\noalign{\smallskip}
            \end{tabular}}
    \caption{The list of the 14 GRBs we collected with $\Pi_{L}$ and redshifts.\\ 
    $\mathrm{I}$: Ref.$~$\citep{Chattopadhyay2022}, ${\mathrm{II}}$: Ref.$~$\citep{GRB061122_pol}, $\mathrm{III}$: Ref.$~$\citep{2021Galax...9...82G}, $\mathrm{IV}$: Ref.$~$\citep{GRB140206A_pol_redshift}.\\ 
    *: 1--6 is the photometric redshift of host galaxy; 7--9, 11--14 is spectral redshift; 10 is the upper limit of redshift.\\ 
    ${a}$: The results obtained from calculations based on photometry data, ${b}$: Ref.$~$\citep{GRB061122_redshift}, ${c}$: Ref.$~$\citep{2022MNRAS.511.1694G}, ${d}$: \url{https://www.mpe.mpg.de/~jcg/grbgen.html} ${e}$: Ref.$~$\citep{GRB140206A_pol_redshift}.}
    \label{tab:Measurements}
\end{table*}

\renewcommand{\dblfloatpagefraction}{.9}
\begin{figure*}[!tbp]
\centering
{\includegraphics[width=3.5in]{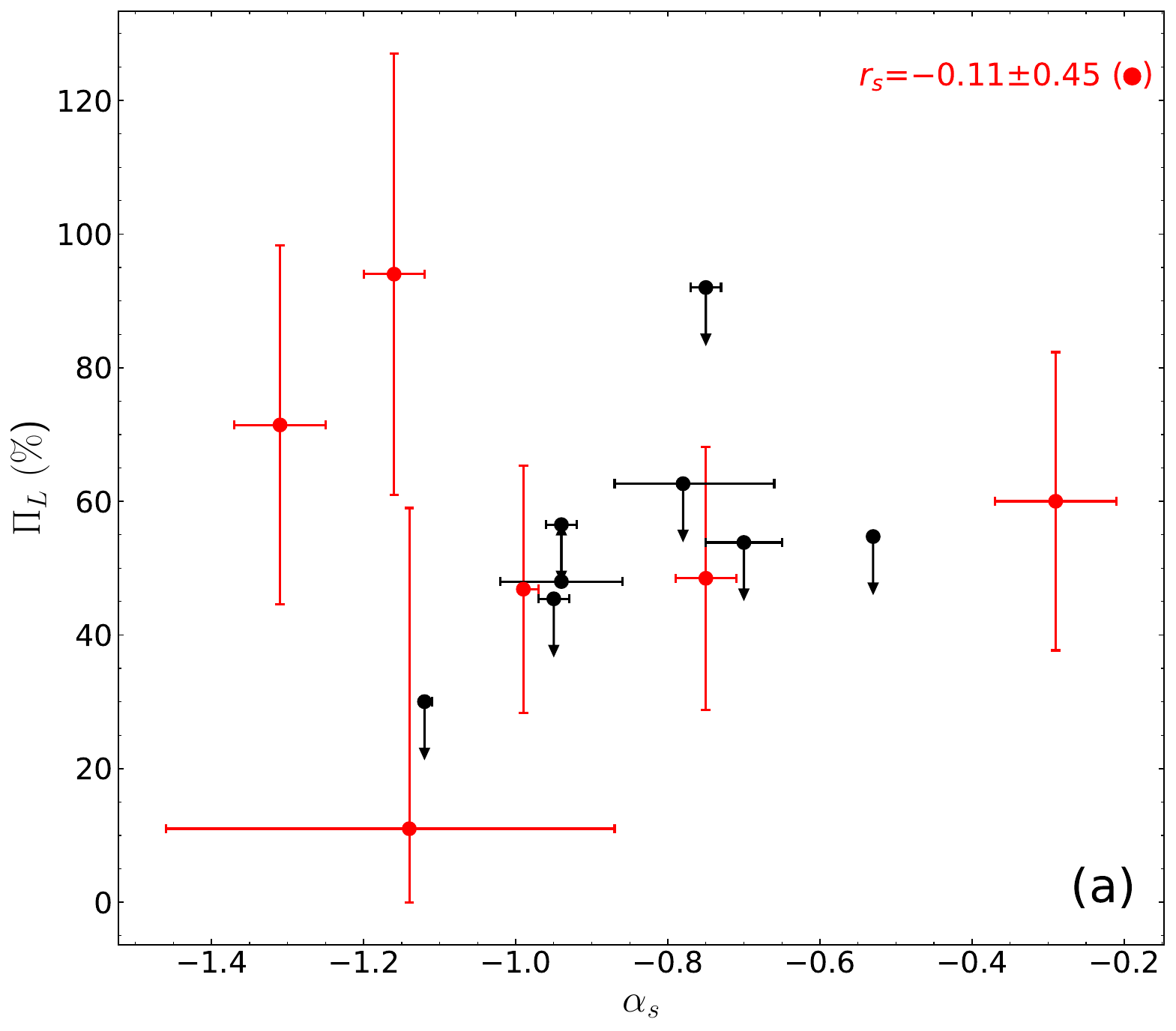}}
{\includegraphics[width=3.5in]{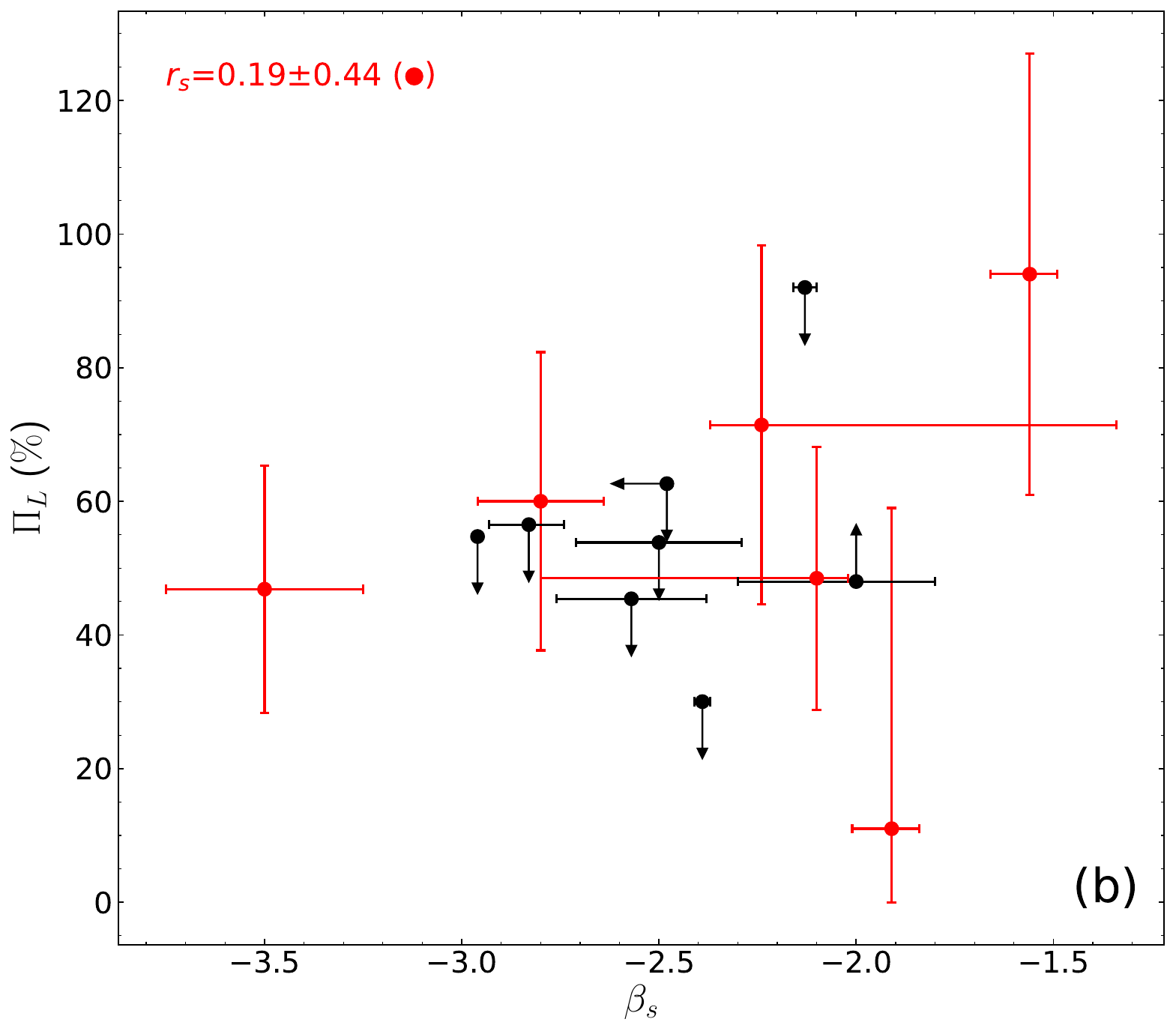}}
{\includegraphics[width=3.5in]{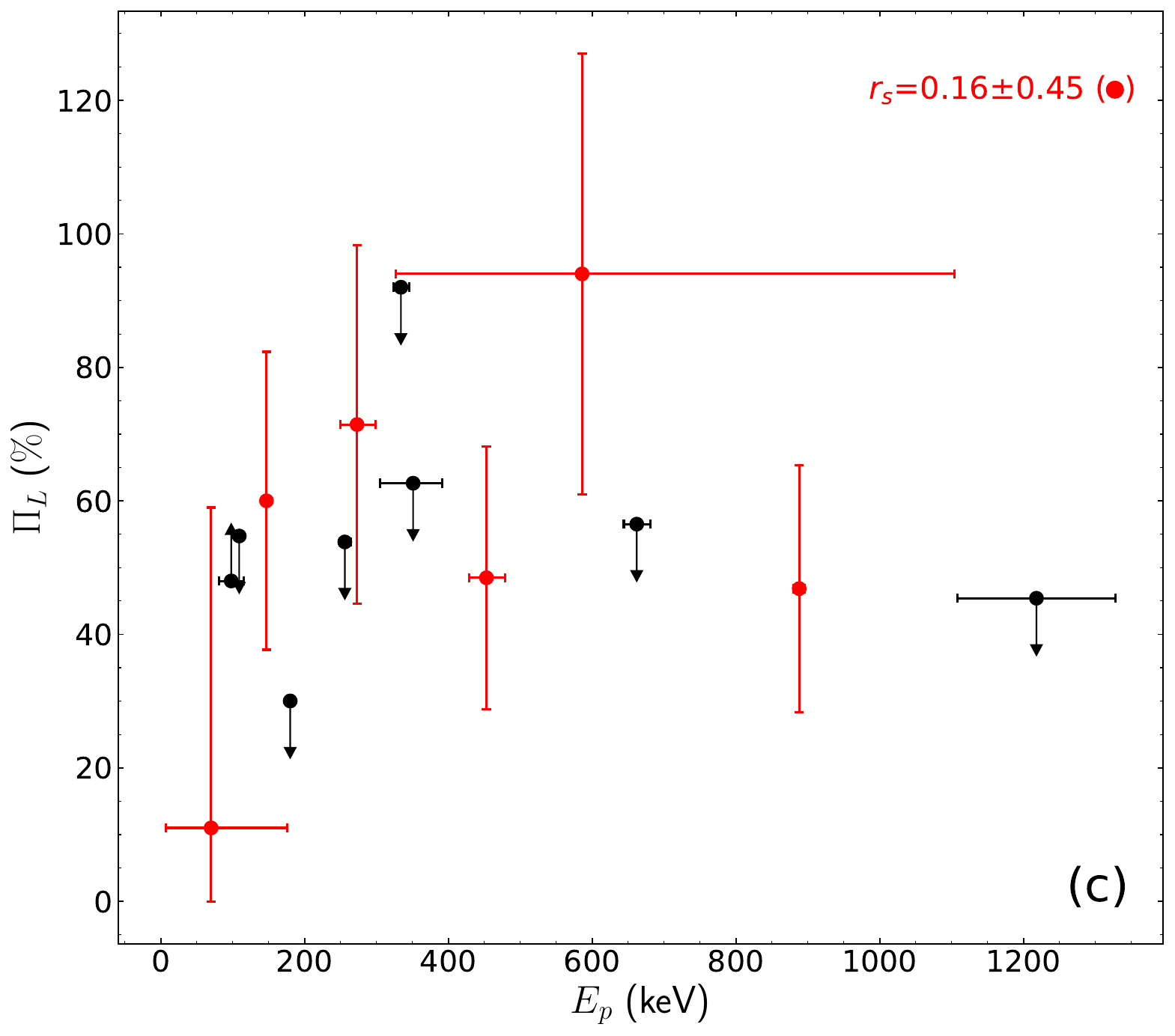}}
{\includegraphics[width=3.5in]{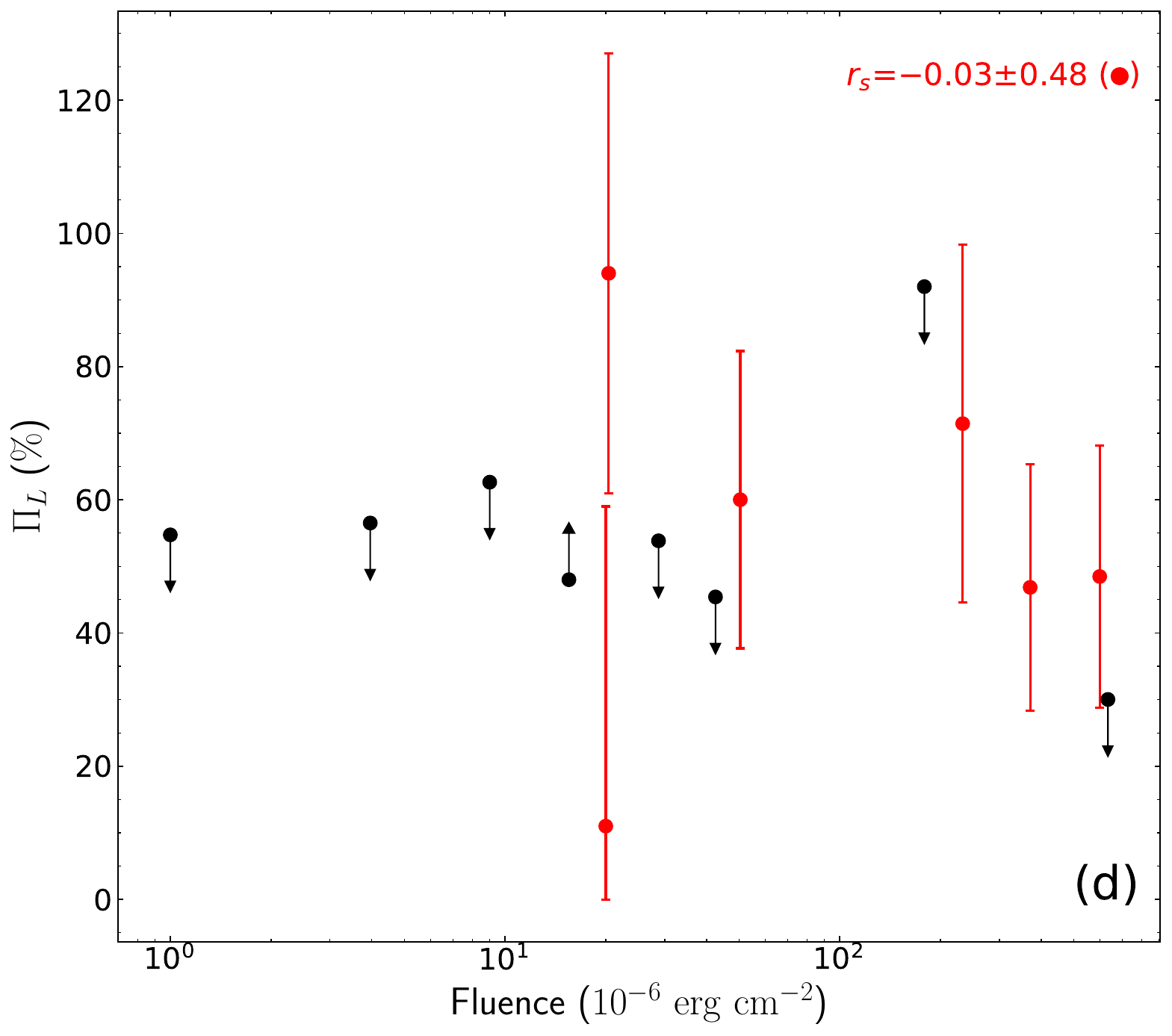}}
\caption{Comparison between $\Pi_{L}$ and the spectral parameters of the GRBs. The red color denotes sources with measured $\Pi_{L}$ whereas black  represents sources with $\Pi_{L}$ constrained as an upper limit or lower limit. (a) $\Pi_{L}$ vs. $\alpha_{s}$ with $r_{s}$=$-0.11\pm$0.45. (b) $\Pi_{L}$ vs. $\beta_{s}$ with $r_{s}$=0.19$\pm$0.44. (c) $\Pi_{L}$ vs. $E_{p}$ with $r_{s}=0.16\pm0.45$. (d) $\Pi_{L}$ vs. Fluence with $r_{s}$=$-0.03\pm$0.48. $r_{s}$ values are exclusively derived from the red dots.}
\label{fig.pol_GRB_correlation}
\end{figure*}

\textbf{\textit{GRBs' $\Pi_{L}$ distribution.}} To begin with, we conducted a thorough investigation of GRBs with available polarimetric observations. 
Based on the recently released catalogs of POLAR and AstroSAT and early observations of INTEGRAL etc.  \citep{2020A&A...644A.124K, Chattopadhyay2022, 2021Galax...9...82G}, we have obtained a sample with 50 GRBs observed by polarimeters in the sub-MeV band. Some of these sources have also been studied individually \citep{GRB061122_pol, GRB140206A_pol_redshift}, and several of them are found in the lists of Ref.$~$\citep{lanmixiang2022} and Ref.$~$\citep{liliang2022nature}. GRB221009A, the BOAT,
has been observed by IXPE, the newly launched polarimeter in the soft X-ray band \citep{2023ApJ...946L..21N}. We have ignored this detection for its meager relevance to the prompt phase. The value of GRB061122A's $\Pi_{L}$ is controversial. The results obtained from Ref.$~$\citep{GRB061122_pol} and Ref.$~$\citep{GRB061122_redshift} are almost incompatible. For it, we used the polarization obtained in Ref.$~$\citep{GRB061122_pol} in the 100--350 keV range. Among the 50 GRBs, there are 28 sources with polarization measurements, 19 with upper limits, and 3 with lower limits. The histogram of $\Pi_{L}$ is shown in Fig.$~$\ref{fig:pol_hist}. Among the 28 measurements, 12 exhibited  $\Pi_{L}$ values below 20$\%$, while 4 sources had $\Pi_{L}$ values above  80$\%$.


Numerical simulations have been conducted in optically thin synchrotron models \citep{Toma2009} or photospheric models with dissipation \citep{2024emission-photosperic}, which are the two most popular models used to explain the origin of prompt radiation. Ref.$~$\citep{Toma2009} investigated the statistical properties of the polarization of GRBs using Monte Carlo simulations, assuming three different jet structures. They discovered that approximately one-third of GRBs exhibit 
$\Pi_{L}$ values clustered between 20\% and 70\%, stemming from synchrotron emission within a globally ordered magnetic field. If several GRBs with $\Pi_{L}>80\%$ are observed, then the Compton Drag model will be favored. In Ref.$~$\citep{2024emission-photosperic}, it is predicted that $\Pi_{L}<4\%$ from the jet core and tends to increase with viewing angle outside of the core, reaching levels as high as 20\%–40\% in photospheric models with dissipation. As shown in Fig.$~$\ref{fig:pol_hist}, the majority of $\Pi_{L}$ values are less than 70\%, with sub-samples clustering below 20\% and several cases exceeding 80\%. Given the current observed number and distribution of $\Pi_{L}$, the controversies persist, and distinguishing between these models remains challenging.

\ \ 

\textbf{\textit{GRBs' spatial distribution.}} The distribution of these 50 GRBs in the sky in Galactic coordinates is shown in Fig.$~$\ref{fig:GRB_distribution}. The dots represent GRBs with observed values of $\Pi_{L}$. The inverted triangles represent GRBs with upper limits while the plus signs represent GRBs with lower limits. The red/blue color indicates GRBs with spectroscopic/photometric redshifts (redshift recognition will be presented in the following paragraph). The relative sizes of the symbols in the figure represent the relative magnitudes of values of $\Pi_{L}$. Despite the limited sample size, the GRBs show no apparent clustering in the sky. 


\ \ 

\textit{\textbf{GRBs' redshifts.}} The presence of ALPs has the potential to alter the initial linear polarization degree of GRB prompt emission. However, the extent of this modification heavily relies on the distance traveled by the GRB within a domain-like structure of ${\bf B}_{\rm IGM}$. Accurate redshift determination is crucial in this context. Hence, we conducted a cross-matching analysis between our sample and the publicly available catalog\footnote{\url{https://www.mpe.mpg.de/~jcg/grbgen.html}}, which provides a subjective collection of the localization of GRBs. Through this analysis, we identified a total of 8 sources that exhibited an overlap. The corresponding sources are displayed in Table \ref{tab:Measurements}, with the redshift values denoted as superscripts `${c}$', `${d}$' or `${e}$'. Among these, 7 sources have spectroscopic redshifts, while 
one has a spectroscopic redshift upper limit.
Regarding the source GRB 190530A, initially available with a spectroscopic redshift upper limit, we acquired its precise spectroscopic redshift from Ref.$~$\citep{2022MNRAS.511.1694G}. For GRB 061122A, we utilized the photometric redshift of its candidate host galaxy in Ref.$~$\citep{GRB061122_redshift}.
For sources lacking documented redshift information, we conducted a search for their candidate host galaxies in the NASA/IPAC Extragalactic Database (NED)\footnote{\url{https://ned.ipac.caltech.edu/}} within an angular separation of $30''$. We assumed that the redshift of a GRB equals that of its candidate host galaxy. Utilizing the photometric data of these candidate host galaxies, we calculated their photometric redshifts using $\texttt{Le PHARE}$\footnote{\url{https://www.cfht.hawaii.edu/~arnouts/LEPHARE/lephare.html}}. Sources lacking sufficient photometric data or exhibiting large fitting errors or significant $\chi^2$ values were excluded from our sample. Additionally, we calculated the offset between the candidate host galaxy and the GRB localization, excluding sources with large offsets. This resulted in 5 GRBs with candidate host galaxies (Table~\ref{tab:Measurements}). 
Our final dataset comprised 14 sources. For a detailed listing of them, along with $\Pi_{L}$ and redshift, refer to Table \ref{tab:Measurements}.


The measured $\Pi_{L}$s of GRBs span a large range, while the mechanisms responsible for the diversity remain inadequately explored. Previous research generally suggests that the inner structure of the GRB itself interacts with 
the host galaxy, potentially influencing this variability. However, the surrounding environment of GRBs and the internal dynamics of the jets are complex and not fully understood, making it challenging to accurately model the emission process. In the subsequent two subsections, we will explore the potential relationships between $\Pi_{L}$ and the spectral characteristics of GRB's prompt emission, as well as those between $\Pi_{L}$ and the properties of GRBs' host galaxies.

\subsection{ $\Pi_{L}$ vs. Spectral Parameters} 

\begin{table*}
\centering
    \resizebox{\textwidth}{!}{
        \begin{tabular}{c c c c c c c c c c}
            \hline
              &GRB & $\Pi_{L}~$($\%$) & $M_{\bigstar}~$($10^{8}~M_{\odot}$) & $M_{\rm gas}~$($10^{8}~M_{\odot}$) & SFR$~$($M_{\odot}$~yr$^{-1}$) & sSFR$~$(Gyr$^{-1}$) &Instrument &Redshift$^{*}$  &Offset\\
            \hline\noalign{\smallskip}
            1 & 200311A  &$<45.41$ & $138.10^{+114.39}_{-114.39}$& $48.05^{+51.67}_{-51.67}$ & $17.56_{-16.72}^{+16.72}$ & $1.27_{-1.60}^{+1.60}$ & AstroSAT$^{\mathrm{I}}$ & $0.0838^{+0.0349}_{-0.0302}$$^{a}$ & $0.072^{\prime}$/8.00kpc\\[5pt]
            2 & 180103A  &$71.43_{-26.84}^{+26.84}$ & $191.99^{+190.94}_{-190.94}$& $52.22^{+82.50}_{-82.50}$ & $84.25_{-50.42}^{+50.42}$ & $4.39_{-5.09}^{+5.09}$ & AstroSAT$^{\mathrm{I}}$ & $0.037^{+0.0063}_{-0.0015}$$^{a}$ & $0.184^{\prime}$/8.75kpc\\[5pt]
            3 & 180427A  &$60.01_{-22.32}^{+22.32}$ & $7.03^{+6.61}_{-6.61}$& $1.89^{+2.88}_{-2.88}$ & $2.87_{-1.74}^{+1.74}$ & $4.08_{-4.57}^{+4.57}$ & AstroSAT$^{\mathrm{I}}$ & $0.0309^{+0.045}_{-0.0309}$$^{a}$ & $0.273^{\prime}$/10.81kpc\\[5pt]
            4 & 200412A  &$<53.84$ & $1611.02^{+1649.92}_{-1649.92}$& $532.48^{+734.13}_{-734.13}$ & $419.58_{-320.89}^{+320.89}$ & $2.61_{-3.33}^{+3.33}$ & AstroSAT$^{\mathrm{I}}$ & $0.1055^{+0.0192}_{-0.0145}$$^{a}$ & $0.131^{\prime}$/18.74kpc\\[5pt]
            5 & 061122A &$11_{-11}^{+48}$ & $96.39^{+65.43}_{-65.43}$& $28.59^{+23.40}_{-23.40}$ & $11.80_{-12.22}^{+12.22}$ & $1.22_{-1.52}^{+1.52}$ & INTEGRAL$^{\mathrm{II}}$  & $1.33_{-0.76}^{+0.77}$$^{b}$ & $0.435^{\prime \prime}$/20.13kpc\\[5pt]
            6 & 200806A  &$<54.73$ & $83.06^{+56.35}_{-56.35}$& $32.10^{+26.47}_{-26.47}$ & $3.64_{-4.85}^{+4.85}$ & $0.44_{-0.65}^{+0.65}$ & AstroSAT$^{\mathrm{I}}$ & $0.1148^{+0.1749}_{-0.1148}$$^{a}$ & $0.234^{\prime}$/36.58kpc\\[5pt]
            \hline\noalign{\smallskip}
        \end{tabular}}
    \caption{ The list of GRBs with candidate host galaxies and their associated properties. $\mathrm{I}$, $\mathrm{II}$, *, a, b is the same as in Table \ref{tab:Measurements}.}
    \label{tab:Host}
\end{table*}

The intense bursts of high-energy gamma rays from GRBs are commonly detected by instruments such as Swift/BAT and Fermi/GBM. The observed nonthermal features of these spectra are typically utilized to distinguish between optical thin synchrotron models \citep{Toma2009} or photospheric models with dissipation \citep{2024emission-photosperic}. However, the issues remain controversial. The observed spectra are frequently fitted using the Band function \citep{Band1993}, which is described by three key parameters: the spectral peak energy $E_{p}$, the low-energy spectral index $\alpha_{s}$, and the high-energy spectral index $\beta_{s}$. Ref.$~$\citep{Toma2009} have found a potential negative correlation between $\Pi_{L}$ and $E_{p}$ arising from synchrotron emission within a globally ordered magnetic field. Here, we also list $\Pi_{L}$ and the corresponding band function parameters for our sample in Table \ref{tab:Measurements}. A significant portion of the parameters is obtained from Ref.$~$\citep{Chattopadhyay2022}, followed by Ref.$~$\citep{Chattopadhyay2019}, Ref.$~$\citep{GRB061122_pol}, and Ref.$~$\citep{GRB140206A_pol_redshift}. Meanwhile, we add the fluence\footnote{\url{https://user-web.icecube.wisc.edu/~grbweb_public/Summary_table.html}} for each GRB to Table \ref{tab:Measurements}. Furthermore, we analyze the potential correlation between $\Pi_{L}$ and these parameters as shown in Fig.$~$\ref{fig.pol_GRB_correlation}, where red dots represent sources with $\Pi_{L}$, and black dots represent sources with only the upper or lower limit of $\Pi_{L}$. Spearman correlation coefficients $r_{s}$ are exclusively derived from the red dots. Fig.$~$\ref{fig.pol_GRB_correlation}(a) illustrates the data points of
$\Pi_{L}$ vs. $\alpha_{s}$. All $\alpha_{s}$ are distributed between $-$1.4 to $-$0.2. About half of them are concentrated around $-$0.85. The Spearman correlation coefficient is $r_{s}=-0.11\pm0.45$. In Fig.$~$\ref{fig.pol_GRB_correlation}(b), the data points of $\Pi_{L}$ vs. $\beta_{s}$ are presented, with $r_{s}=0.19\pm0.44$. The distribution of all $\beta_{s}$ values falls within the range of $-$3.5 to $-$1.5. Most of them are concentrated around $-$2.5. Fig.$~$\ref{fig.pol_GRB_correlation}(c) shows the data points of $\Pi_{L}$ vs. $E_{p}$, with $r_{s}=0.16\pm0.45$. These data points are scattered within the range of 50 keV to 1250 keV. Most of them are less than 500 keV. The Band function has a smoothly broken power law with spectral peak energy at $E_{p}$$\simeq$210 keV in $\nu$$F_{\nu}$ space, featuring asymptotic power-law photon indices below ($\alpha_{s}$$\simeq -0.8$) and above ($\beta_{s}$$\simeq -2.5$) the break energy \citep{liliang2023, Toma2009}. Our $\alpha_{s}$ and $\beta_{s}$ match the features, but $E_{p}$ is not evident. In Fig.$~$\ref{fig.pol_GRB_correlation}(d), the data points of $\Pi_{L}$ vs. Fluence are shown, with $r_{s}=-0.03\pm0.48$. They all fall between 1$\times$10$^{-6}$\,erg\,cm$^{-2}$ and 6.5$\times$10$^{-4}$\,erg\,cm$^{-2}$. The data points are widely dispersed, with all correlation coefficient absolute values obtained being less than 0.3. Meanwhile, the errors associated with these coefficients exceed the coefficients themselves. Consequently, it is not possible to determine whether there exists a correlation between $\Pi_{L}$ and the spectral parameters, limiting our ability to confirm or rule out the correlation predicted by Ref.$~$\citep{Toma2009}. 

The inclusion of photon-ALP mixing further complicates this issue, particularly when considering mixing within the inner structure of GRBs. However, the precise location of photon-ALP mixing within the source remains elusive. Some studies suggest that mixing occurs in the external shock \citep{galanti_PRL} with ${\bf B}_{\rm jet}$ around $\mathcal{O}(1)$ G, while others explore the possibility of mixing in the internal shock with ${\bf B}_{\rm jet}$ ranging from $10^{6}$ G \citep{GRBpol-jet} to $10^{9}$ G \citep{GRBpol-IGM}. It is imperative to thoroughly investigate potential deviations in both the early-time evolution pattern of $\Pi_{L}$ and the spectral shape from predictions made by traditional emission models. However, it is important to note that the detected $\Pi_{L}$ values, as depicted in Fig.$~$\ref{fig:pol_hist}, are averaged over both energy and time. In the future, with the discovery of closer sources and the minimization of the influence of $\bf B$$_{\rm IGM}$, coupled with time-resolved high-sensitivity observations, there exists the possibility of more accurately probing the impact of the magnetic field within the source. In the current paper, our focus will be on assessing the influence of $\bf B$$_{\rm IGM}$ on the initial  $\Pi_{L}$.

\subsection{$\Pi_{L}$ vs. Host Galaxies' Properties}

\renewcommand{\dblfloatpagefraction}{.9}
\begin{figure*}[!tbp]
\centering
{\includegraphics[width=3.5in]{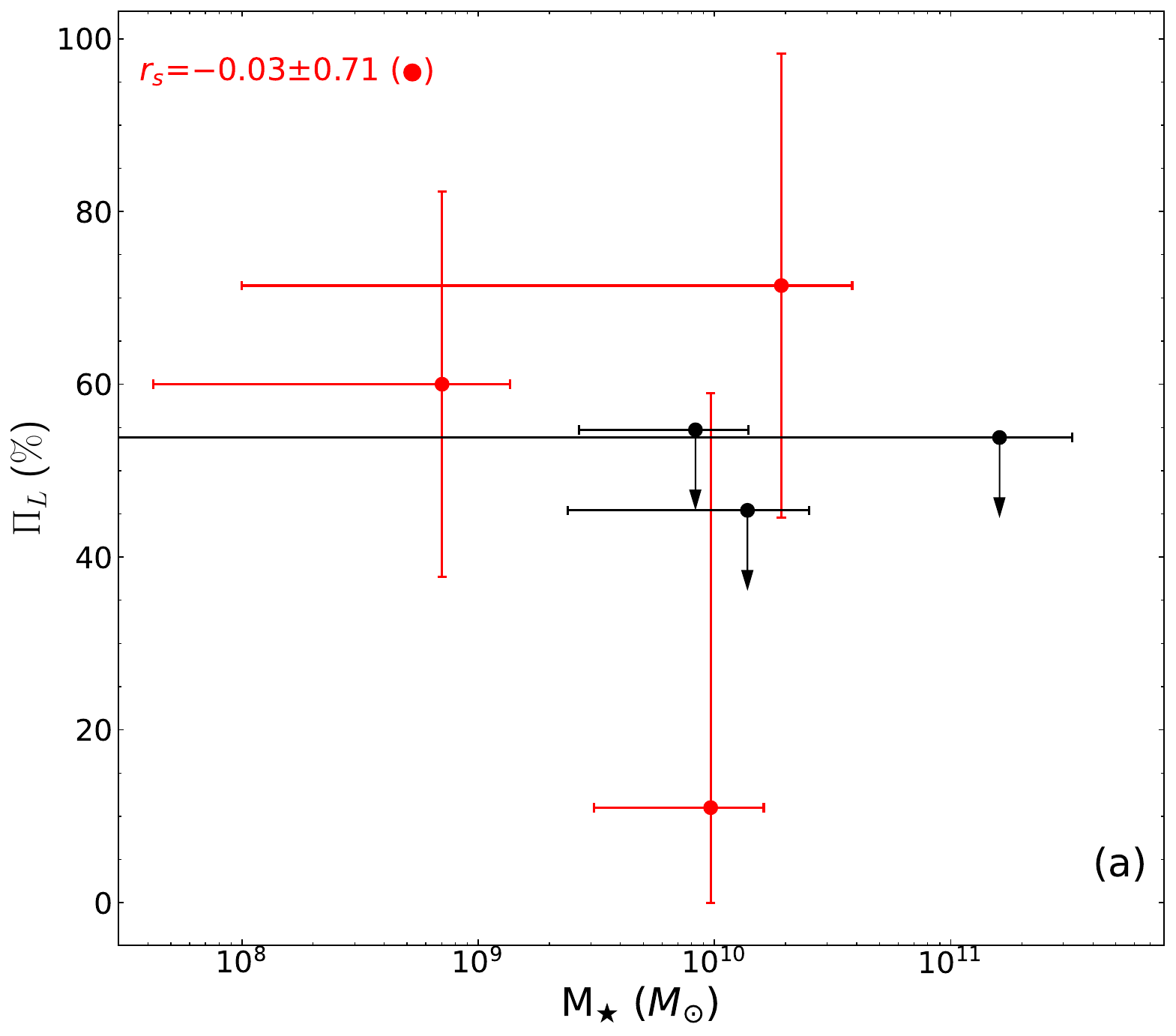}}
{\includegraphics[width=3.5in]{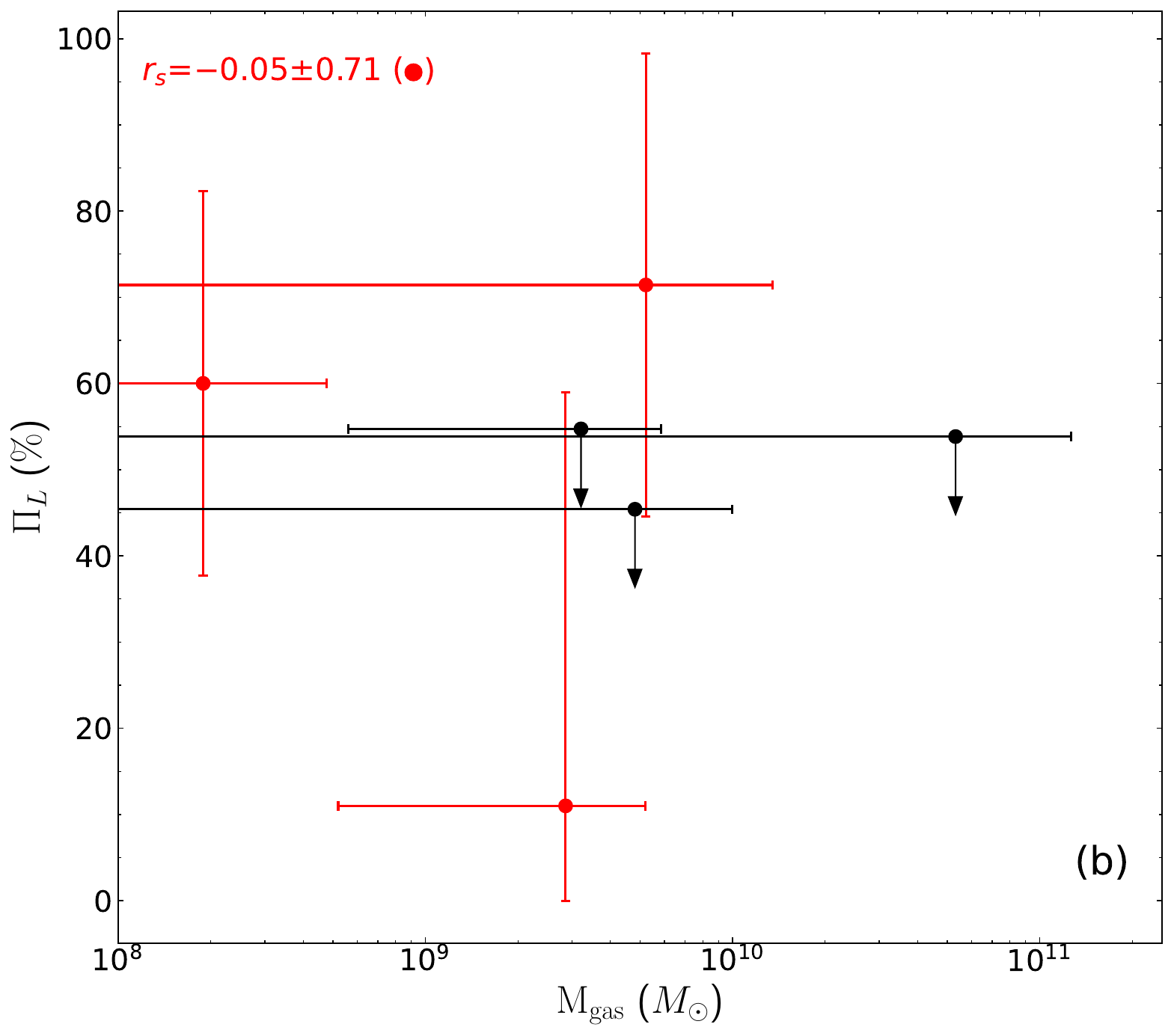}}
{\includegraphics[width=3.5in]{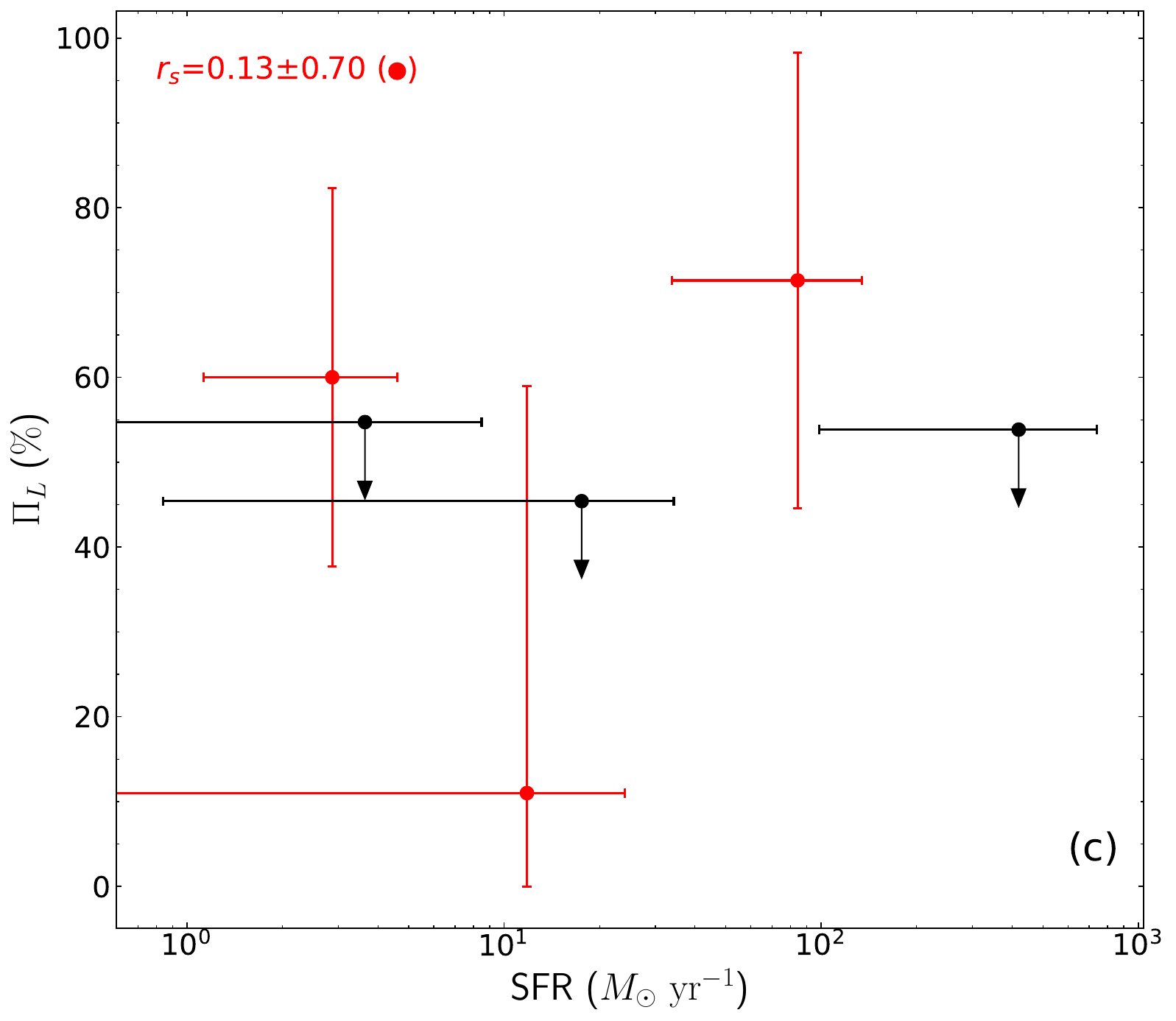}}
{\includegraphics[width=3.5in]{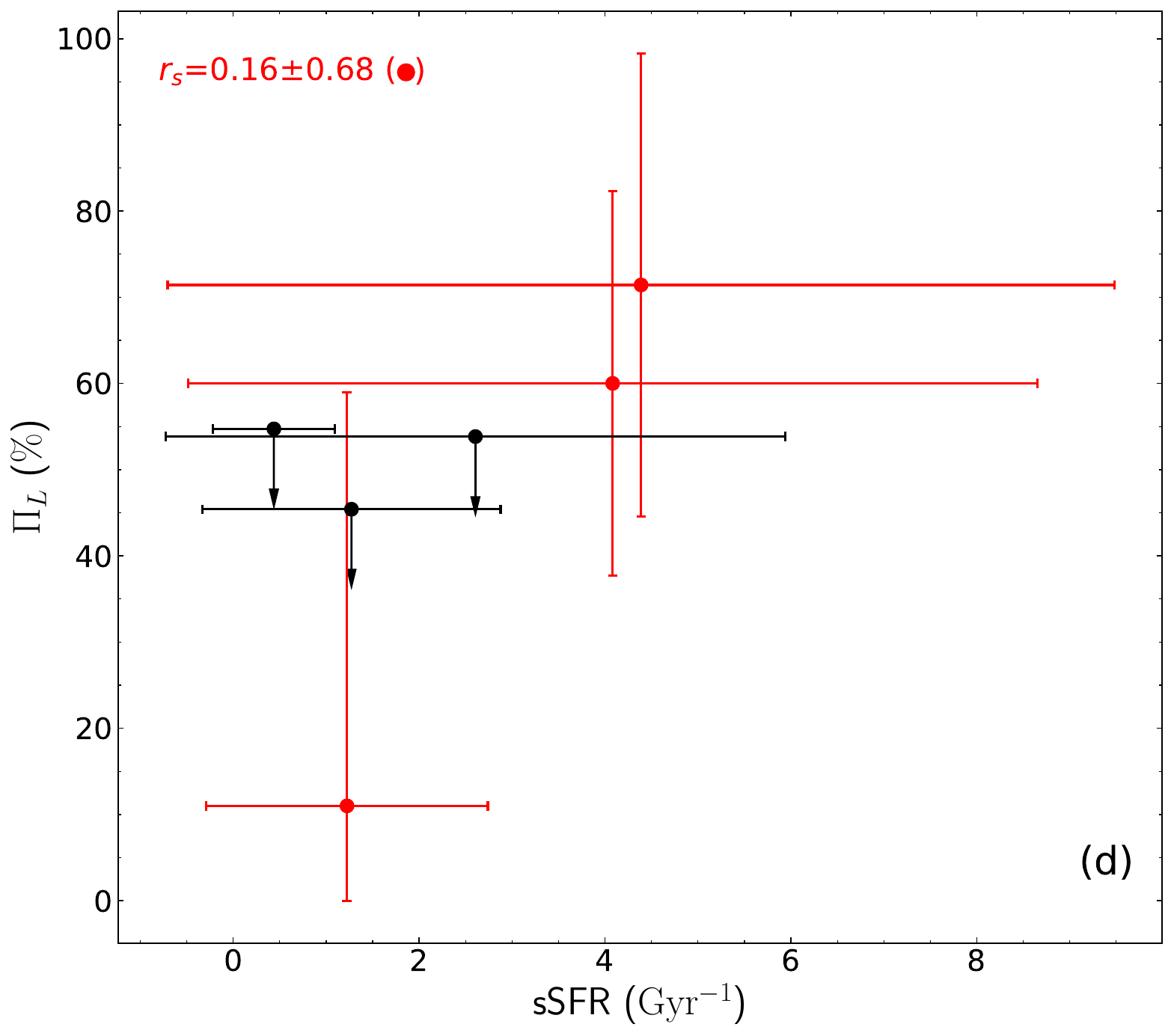}}
\caption{Comparison between $\Pi_{L}$ and the properties of the GRB candidate host galaxies. Red color denotes sources with measured $\Pi_{L}$, whereas black color represents sources with $\Pi_{L}$ constrained as an upper limit.  (a) $\Pi_{L}$ vs. M$_{\bigstar}$ with $r_{s}$=$-0.03\pm$0.71. (b) $\Pi_{L}$ vs. M$_{\rm gas}$ with  $r_{s}$=$-0.05\pm$0.71. (c) $\Pi_{L}$ vs. SFR with $r_{s}$=0.13$\pm$0.70. (d) $\Pi_{L}$ vs. sSFR with $r_{s}$=0.16$\pm$0.68.}
\label{fig.pol_host_correlation}
\end{figure*}

The intrinsic characteristics of a GRB itself may contribute to the variability of $\Pi_{L}$, while the properties of the GRB host galaxies could also influence $\Pi_{L}$. To explore this possibility, we analyzed the photometric data obtained from six candidate host galaxies of GRBs, as listed in Table \ref{tab:Host}. Utilizing Spectral Energy Distribution (SED) fitting via the CIGALE code, we derived various galactic properties, including the stellar mass (M$_{\bigstar}$), gas mass (M$_{\rm gas}$), star formation rate (SFR), specific star formation rate (sSFR) for these candidate host galaxies. The results of this analysis are presented in Table \ref{tab:Host}. We then investigated the potential correlation between the measured $\Pi_{L}$ and these galactic properties. The results are presented in Fig.$~$\ref{fig.pol_host_correlation}. 
Red data points denote sources with $\Pi_{L}$, whereas black data points represent sources with an upper limit of $\Pi_{L}$. Correlation calculations exclusively utilized the red data points.

Fig.$~$\ref{fig.pol_host_correlation}(a) displays comparison between the measured $\Pi_{L}$ and M$_{\bigstar}$. The latter of the candidate host galaxies for GRBs ranges between $6\times10^{8}$ $M_{\odot}$ and $2\times10^{11}$ $M_{\odot}$, with approximately half centered around $10^{10}$\,$M_{\odot}$, smaller than that in the Milky Way. The Spearman correlation coefficient $r_{s}=-0.03\pm0.71$. Extinction ($A_{v}$) for candidate host galaxies was also calculated, with only one showing a small extinction close to 1, while the rest exhibit zero extinction. This low or zero extinction suggests a limited amount of dust present in these galaxies. Fig.$~$\ref{fig.pol_host_correlation}(b) displays comparison between the measured $\Pi_{L}$ and M$_{\rm gas}$. All M$_{\rm gas}$ of the candidate host galaxies for GRBs span from $10^{8}$\,$M_{\odot}$ to $10^{11}$\,$M_{\odot}$, with a majority clustered around $4\times10^{9}$\,$M_{\odot}$. Distribution of M$_{\rm gas}$ resembles that of M$_{\bigstar}$ and the Spearman correlation coefficient is $r_{s}=-0.05\pm0.71$. Fig.$~$\ref{fig.pol_host_correlation}(c)/(d) demonstrates a comparison between the measured $\Pi_{L}$ and SFR/sSFR in the candidate host galaxies. The Spearman correlation coefficient is $r_{s}=0.13\pm0.70$/$r_{s}=0.16\pm0.68$. The majority of SFR values are less than tens of $M_{\odot}$\rm yr$^{-1}$ with a single exception displaying a value of several hundred $M_{\odot}$ yr$^{-1}$, akin to starburst galaxies. All sSFRs are under 5 $\rm Gyr^{-1}$. The absolute values of the above Spearman correlation coefficients are all less than $0.2$ based on the limited sample and the significant uncertainties. We are unable to draw any conclusions about the correlation between the measured $\Pi_{L}$ and these galactic properties. However, we can infer that with diversity in the properties of candidate host galaxies, the $\Pi_{L}$ of emission from GRB's prompt phase could become more diverse, especially when considering the photon-ALP mixing in the host magnetic fields ${\bf B}_{\rm host}$. Hence, we do not consider the influence of ${\bf B}_{\rm jet}$ or ${\bf B}_{\rm host}$, our focus will be on assessing the influence of $\bf B$$_{\rm IGM}$ on the initial linear polarization degree ($\Pi_{L0}$), which refers to the polarization state of photons emitted by GRBs as they enter the intergalactic medium.

\subsection{$\Pi_{L}$ vs.  GRBs' Redshifts}

The correlation between spectral slope and redshift in the observed $\gamma$-ray spectra of blazars has been considered as an indication of photon-ALP mixing in ${\bf B}_{\rm IGM}$ \citep{2009AngelisAGNz-first,2020Gal-blazar-z,2023Dong-AGN-z}. It is worth exploring whether the same pattern may also exist among the data points of the measured $\Pi_{L}$ and redshifts, which are shown in Fig.$~$\ref{fig:pol_Zhost}.
The Spearman correlation coefficient obtained from sources with confirmed polarization values is $r_{s}$~$=$~$-0.18\pm0.45$, indicating a lack of statistical behavior in describing the evolution of measured $\Pi_{L}$ over redshift. In the context of photon-ALP mixing in ${\bf B}_{\rm IGM}$, we could still utilize the distribution of the measured $\Pi_{L}$ with redshifts to impose constraints on the mixing, or to explore the allowing parameter space of ALPs and ${\bf B}_{\rm IGM}$ that could allow us to safely ignore the propagation effects on the $\Pi_{L_0}$.


\section{photon-ALP mixing}
\label{Sec_mixing}

\subsection{Basic model}
\label{means}
This section will illustrate the basic framework related to photon-ALP mixing in external magnetic fields. The Lagrangian of photon-ALP coupling is generally described by the following equation:
\begin{equation}
{\cal L}_{a\gamma}=-\frac{1}{4} \,\gag
F_{\mu\nu}\tilde{F}^{\mu\nu}a=\gag \, {\bf E}\cdot{\bf B}\,a~,
\end{equation}
where $F_{\mu\nu}$ and $\tilde{F}^{\mu\nu}$ are the electromagnetic field tensor and its dual, respectively. We initiate our analysis with a photon beam propagating along the $x_{3}$-direction. The beam propagation equation can be written in the following Schr\"odinger-like one
\begin{equation}
\label{we} 
\left(i \, \frac{d}{d x_{3}} + E +  {\cal M} \right)  \left(\begin{array}{c}A_{x_{1}} (x_{3}) \\ A_{x_{2}} (x_{3}) \\ a (x_{3}) \end{array}\right) = 0~,
\end{equation}
where $E$ is the energy of photon-ALP beam, ${\cal M}$ is the photon-ALP mixing matrix, $A_{x_{1}} (x_{3})$ and $A_{x_{2}} (x_{3})$ are the photon linear polarization amplitudes along the $x_{1}$ and $x_{2}$ axis, respectively, and $a (x_{3})$ is the ALP amplitude. ALPs and photon couple in the transverse magnetic field ${\bf B}_T$, namely magnetic field component in the $x_{1}$-$x_{2}$ plane. 

The angle formed by ${\bf B}_T$ and $x_{2}$-axis is $\psi$. When $\psi$ is 0, the mixing matrix in the homogeneous magnetic field can be simplified to \citep{GRBpol-IGM}
\begin{equation}
{\cal M}^{(0)} =   \left(\begin{array}{ccc}
\Delta_{ \perp}  & 0 & 0 \\
0 &  \Delta_{ \parallel}  & \Delta_{a \gamma}  \\
0 & \Delta_{a \gamma} & \Delta_a 
\end{array}\right)~,
\label{eq:massgen}
\end{equation}
where
\begin{equation}
\begin{split}
\Delta_{a\gamma} &\equiv \frac{1}{2} g_{a\gamma} B_{T} \\
&\simeq 1.5 \times 10^{-2} \frac{g_{a\gamma}}{10^{-11} \rm GeV^{-1}} \frac{B_{T}}{\rm nG} {\rm Mpc}^{-1},  \\
\Delta_a &\equiv - \frac{m_{a}^2}{2E} \\
&\simeq -7.8 \times 10^{-3} \left(\frac{m_a}{10^{-13} \, {\rm eV}}\right)^2 \left(\frac{E}{{10^2 \, \rm keV}} \right)^{-1} {\rm Mpc}^{-1}~, \\ 
\Delta_{\rm pl} &\equiv -\frac{\omega^2_{\rm pl}}{2E} \\
&\simeq - 1.1 \times10^{-4} \left(\frac{E}{{10^2\, \rm keV}}\right)^{-1} \frac{n_e}{10^{-7} \, {\rm cm}^{-3}} {\rm Mpc}^{-1}~,\\  
\Delta_{\rm QED} &\equiv \frac{\alpha E}{45 \pi} \left(\frac{B_T}{B_{\rm cr}} \right)^2 \\
&\simeq  4.1 \times 10^{-16}\frac{E}{{10^2\, \rm keV}} \left(\frac{B_T}{10^{-9}\,\rm G}\right)^2 {\rm Mpc}^{-1}~,\\ 
\Delta_\parallel &\equiv \Delta_{\rm pl} + 3.5 \, \Delta_{\rm QED},\  \Delta_\perp \equiv \Delta_{\rm pl} + 2 \, \Delta_{\rm QED}\ .
\end{split}
\end{equation}

\begin{figure}[htp]
\begin{minipage}[t]{0.99\linewidth}
\centering
\includegraphics[width=1.0\textwidth]{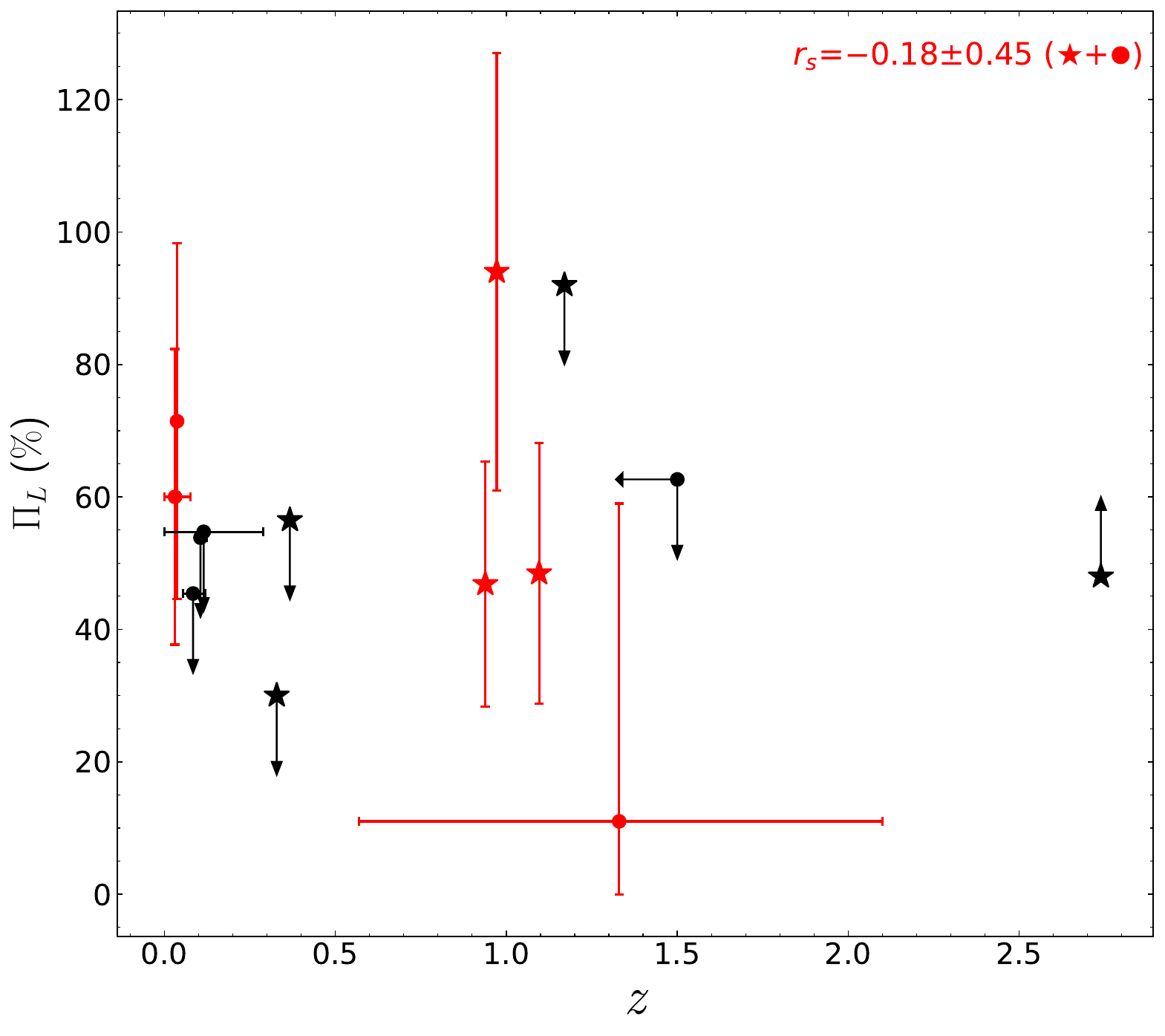}
\caption{Comparison between $\Pi_{L}$ and redshifts of the GRBs. The pentagon denotes sources with spectroscopic redshift, while the dot represents sources without spectroscopic redshift, as well as sources with only upper limits on their redshift. The correlation coefficients are $r_{s}=-0.18\pm0.45$ for the red ones.}
\label{fig:pol_Zhost}
\end{minipage}
\end{figure}

After traveling a distance of $d$, the probability of the photon's initial linear polarization along the $x_{2}$-axis transforming into ALPs is 
\begin{equation}
\label{eq_prob}
P^{(0)}_{\gamma \rightarrow a} ={\rm sin}^2 2 \theta \  {\rm sin}^2\left( \frac{\Delta_{\rm osc} \, d}{2} \right)~, 
\end{equation}
where the oscillation wave number is 
\begin{equation}
\label{eq_osc}
{\Delta}_{\rm osc} = \left[\left( \Delta_a - \Delta_{\parallel} \right)^2 + 4 \Delta_{a \gamma}^2 \right]^{1/2}~
\end{equation}
and the mixing angle is 
\begin{equation}
\label{eq_mixang}
\theta = \frac{1}{2} \arctan \left(\frac{ 2 \Delta_{a \gamma}}{\Delta_{\parallel}-\Delta_a} \right). 
\end{equation}
When ${\rm sin}^2 2 \theta\simeq 1$, that is $\theta \simeq \pi/4$, photon-ALP oscillations reach their maximum. This condition is then determined by
\begin{equation}
\label{eq_comp}
 |\Delta_{\parallel} -\Delta_a| \ll 2 \Delta_{a \gamma} .
\end{equation}
If the $\Delta_{\rm QED}$ term could be ignored where ${{\bf B}_{T}}\ll {\bf B}_{cr}$$\simeq 4.41\times 10^{13} $ G, $\Delta_\parallel \simeq \Delta_\perp \simeq \Delta_{\rm pl}$. We then can define a low energy limit as 
\begin{equation}
\begin{split}
\label{eq_EL}
E_L &\equiv  \frac{E \, |\Delta_a- \Delta_{\rm pl}|}{2 \, \Delta_{a \gamma}} \\
&\simeq \frac{25 \, | m_a^2 - {\omega}_{\rm pl}^2|}{(10^{-13}{\rm eV})^2} \left( \frac{{\rm nG}}{B_T} \right) \left( \frac{10^{-11}\rm GeV^{-1}}{g_{a \gamma}} \right) {\rm keV}.
\end{split}
\end{equation}
If ${\bf B}_{T}$ approaches $ {\bf B}_{cr}$, a high energy limit defined as 
\begin{equation}
\begin{split}
E_H &\equiv \frac{90 \pi \,g_{a \gamma} \, B^2_{\rm cr}}{7 \alpha \,B_T} \\
&\simeq 2.1\times 10^{15} \left(\frac{\rm{nG}}{B_T}\right)  \left(\frac{g_{a \gamma}}{10^{-11}\rm GeV^{-1}}\right) {\rm keV}~.
\end{split}
\end{equation}
When $ E_L\ll E \ll E_H$, photon-ALP oscillations reach their maximum (generally called strong mixing), and the conversion probability becomes energy-independent. The above-associated plasma frequency
\begin{equation}
\label{eq_plasma}
\omega_{\rm pl} = \sqrt{4 \pi \alpha n_e/m_e } = 3.71 \times 10^{-14} \sqrt{\frac{n_e}{\rm cm^{3}}} \  \rm keV\ ,
\end{equation}
where $n_e$ is the electron density in the medium. 

Moreover, the initial polarization state of the photon is expected to change as a result of its interaction with spin-zero ALPs. Typically, we employ the Stokes parameters to characterize the polarization state, defined as follows: 
\begin{eqnarray}
\begin{split}
\rm{I}&=\left \langle A^{*}_{x_{2}}(x_{3})A_{x_{2}}(x_{3})\right \rangle +\left \langle A^{*}_{x_{1}}(x_{3})A_{x_{1}}(x_{3})\right \rangle , \\
\rm{Q}&=\left \langle A^{*}_{x_{2}}(x_{3})A_{x_{2}}(x_{3})\right \rangle - \left \langle A^{*}_{x_{1}}(x_{3})A_{x_{1}}(x_{3})\right \rangle , \\
\rm{U}&=2 Re\left \langle A^*_{x_{2}}(x_{3})A_{x_{1}}(x_{3})\right \rangle , \\
\rm{V}&=2 Im\left \langle A^*_{x_{2}}(x_{3})A_{x_{1}}(x_{3})\right \rangle ,
\end{split}
\end{eqnarray}
where $A^{*}_{x_{1}}(x_{3})$ ($A^{*}_{x_{2}}(x_{3})$) is the conjugation of $A_{x_{1}}(x_{3})$ ($A_{x_{2}}(x_{3})$).
The linear polarization degree $\Pi_{L}$ can be expressed by stokes parameters as
\begin{equation}
\Pi_{L} = \frac{\sqrt{\rm{Q}^2+\rm{U}^2}}{\rm{I}}.
\end{equation}

In the literature, it is more convenient to use the polarization density matrix and its transfer matrix for numerical calculation. The polarization density matrix is defined as  \citep{GRBpol-IGM}
\begin{equation}
\rho (x_{3}) = \left(\begin{array}{c}A_{x_{1}} (x_{3}) \\ A_{x_{2}} (x_{3}) \\ a (x_{3})
\end{array}\right)
\otimes \left(\begin{array}{c}A_{x_{1}} (x_{3})\  A_{x_{2}} (x_{3})\ a (x_{3})\end{array}\right)^{*},
\end{equation}
which obeys the Liouville-Von Neumann equation
\begin{equation}
\label{vne}
i \frac{d \rho}{d x_{3}} = [\rho, {\cal M}].
\end{equation}
We use  $T(x_{3},x_{3}^{0})$ to represent the transfer matrix. Then $T(x_{3}^{0},x_{3}^{0}) = 1$ is the initial condition for solution of Eq.~(\ref{we}). The solution of Eq.~(\ref{vne}) can be expressed by the transfer equation as follows \citep{GRBpol-IGM}
\begin{equation}
\label{vnea}
\rho (x_{3}) = T(x_{3},x_{3}^{0}) \, \rho (x_{3}^{0}) \, T^{\dagger}(x_{3},x_{3}^{0})~.
\end{equation}

The comprehensive procedure for solving $\rho(x_{3})$ is not outlined here. For a detailed derivation of $\rho(x_{3})$, refer to Section 3 of Ref.$~$\citep{GRBpol-IGM}, specifically equations (9) to (42). The detailed derivation of the expressions of stokes parameters in terms of the polarization density matrix is also presented in 
Ref.$~$\citep{Ganguly12}.
We give the final expression of the 2 × 2 photon polarization density matrix (i.e. the 1--2 block of the density matrix for the photon-ALP system) here:  
\begin{equation}
\rho_{\gamma}=\frac{1}{2}\left(\begin{matrix}I+Q&U-iV\\ U+iV&I-Q\end{matrix}\right)~,
\end{equation}
and the linear polarization degree is then expressed as follows:
\begin{equation}
{\Pi}_L  \equiv \frac{\sqrt{Q^2 + U^2}}{I} =
 \frac{ \sqrt{\left( \rho_{11} - \rho_{22} \right)^2 + \left( \rho_{12} + \rho_{21} \right)^2 }}{ \rho_{11} + \rho_{22} }~.
 \end{equation}

Through this paper, we utilize the gammaALPs tool, publicly accessible at \url{https://gammaalps.readthedocs.io/en/latest/}, to numerically solve the transfer equation for photon-ALP mixing. This code is specifically designed to handle complex scenarios involving diverse initial states and arbitrary orientations of ${\bf B}_T$ with an angle $\psi$ relative to the $x_{2}$-axis within different magnetic configurations. The photon-ALP beam, during its propagation from the source to Earth, can traverse various magnetic fields including those associated with the source (eg $\bf B$$_{\rm jet}$), the host galaxy ($\bf B$$_{\rm host}$), the intergalactic medium ($\bf B$$_{\rm IGM}$), and the Milky Way ($\bf B$$_{\rm MW}$). If located within a galaxy cluster, the photon-ALP beam would also traverse the magnetic field of the intra-cluster medium ($\bf B$$_{\rm ICM}$). 

\subsection{Basic parameters}
In Sec.$~$\ref{Sec_Statistical}, we have explored the statistical characteristics of $\Pi_{L}$. Predictions for $\Pi_{L}$ from numerical simulations of popular emission mechanisms during the prompt phase of GRBs range from as low as $4\%$ to values exceeding $80\%$. Uncertainties regarding the jet structure and the diverse properties of host galaxies further complicate this issue, particularly in the context of photon-ALP mixing within ${\bf B}_{\rm jet}$ and/or ${\bf B}_{\rm host}$. Therefore, we consider the polarization state of photons emitted by GRBs as they enter the intergalactic medium as our initial $\Pi_{L_0}$, which ranges from unpolarized case ($\Pi_{L_0}=0$) to fully linearly polarized case ($\Pi_{L_0}=100\%=1$), focusing our analysis on the influence of ${\bf B}_{\rm IGM}$ and ${\bf B}_{\rm MW}$ on the $\Pi_{L0}$ for GRBs across varied redshifts.

For $\bf B$$_{\rm IGM}$, our current understanding is very limited. Observations of Faraday rotation in both radio and optical wavelengths from distant astronomical sources have revealed the presence of weak yet coherent magnetic fields in extragalactic space. However, the origin, intensity, and configuration of these magnetic fields remain subjects of ongoing research and debate \citep{IGM2013-review}. The dynamical amplification of primordial seed fields, whether generated from the early universe or by motions of the plasma in (proto) galaxies, could, in principle, lead to the establishment of such a large-scale magnetic field. 
However, there is considerable uncertainty and complexity in the physical processes involved, making theoretical predictions highly uncertain. Meanwhile, a large region of the strength of $\bf B$$_{\rm IGM}$ has been obtained in observations. The lower limits for $\bf B$$_{\rm IGM}$ can be obtained based on the electromagnetic cascade generated by the deflection of electron-positron pairs by the $\bf B$$_{\rm IGM}$ with the value around the order of $10^{-14}$ G \citep{Neronov2010-IGM-low, 2023Aharonian-IGM-low, 2023huang-IGM-low, 2024Meyer-IGM-low}. Faraday rotation measurements have imposed upper limits on $\bf B$$_{\rm IGM}$ around $\mathcal{O}(1)$ nG for coherence lengths of about $\mathcal{O}(1)$ Mpc, while the bound decreases with the increasing correlation length \citep{2016Pshirkov-IGM-up}. The above limits on the strength of $\bf B$$_{\rm IGM}$ are deduced from limited observational data. Precise measurements remain challenging due to the diffuse and low-density nature of the intergalactic medium. In the scenario of photon-ALP mixing, the $\bf B$$_{\rm IGM}$ approaching the current upper limit are typically assumed to investigate the maximum potential effects.
In Ref.$~$\citep{2013PhRvD..87c5027M}, numerical calculations showed that the impact of photon-ALP mixing increases with the intensity of $\bf B$$_{\rm IGM}$ among the strong mixing energy region. As indicated by Eq.~(\ref{eq_osc}), when $E_L \ll E \ll E_H$, photon-ALP mixing becomes energy-independent, with the mixing term $\Delta_{a\gamma} \equiv \frac{1}{2} g_{a\gamma} {\bf B}_{\rm IGM}$ serving as the key factor reflecting the strength of the induced effects. Any constraints on the mixing effect can, in turn, be regarded as constraints on $g_{a\gamma} {\bf B}_{\rm IGM}$. Here, we adopt that $\bf B$$_{\rm IGM}$$~$=$~$1$~$ nG and coherence length $L_{\rm coh}$$~$=$~$$1$$~$ Mpc at $z~$=$~$0. For $n_e$ in the intergalactic medium, we use the typical values $10^{-7} \rm cm ^{-3}$ derived from the baryon density \citep{Jarosik2011_IGM_n_e} and the corresponding plasma frequency is $1.17\times 10^{-14}$ eV as calculated with Eq.~(\ref{eq_plasma}). Regarding the configuration of ${\bf B}_{\rm IGM}$, a conventional domain-like model is commonly utilized. In this model, the magnetic field strength remains constant within each domain, while the direction changes randomly from one domain to another. Ref.$~$\citep{GRBpol-IQUV} compare this traditional domain-like model with another domain-like model, wherein the magnetic field is helical within each domain. They argue that these two models show a similar asymptotic behavior. Here, we follow the steps in Ref.$~$\citep{gammaALPs2022} and adopt the conventional domain-like model. To achieve this, we treat $\psi$ 
as a random variable in the range of 0 to 2$\pi$ for each domain with size $L_{\rm coh}$ and constant ${\bf B}_{\rm IGM}$. As the photon beam travels along the $x_{3}$-axis from the source to the observer, it will experience a series of random magnetic field domains. One set of $\psi$ values means one realization of the random magnetic field configurations. 

 
For $\bf B$$_{\rm WM}$, we have a more comprehensive understanding of its 
morphology. We consider the model developed by Jansson and Farrar to be a well-established one \citep{galanti_PRL, 2012ApJ...757...14J, 2012ApJ...761L..11J, 2016JCAP...05..056B}. 
The photons of GRBs from different lines of sight undergo different parts of the  magnetic fields during their propagation within the Milky Way. If the mixing in the Galactic magnetic field is significant, it is anticipated that the effects of mixing would vary depending on Galactic positions of the GRBs. However, as we examine later, mixing in  $\bf B$$_{\rm WM}$ is negligible within the parameter space of ALP and the energy band we are exploring.


Another crucial set of parameter is $m_a$ and $g_{a\gamma}$. The most reliable constraints on ALPs come from the none detection of ALPs from the Sun derived by the CERN Axion Solar Telescope (CAST), yielding an upper limit for $g_{a\gamma}$ as $6.6 \times 10^{-11}$ GeV$^{-1}$ for $m_{a} \lesssim 0.02$ eV \citep{CAST_limit}. For $m_a$ values considerably exceeding the plasma frequency $\omega_{\rm pl}$, the mass term will predominate over the mixing term. Conversely, for $m_a$ values much less than $\omega_{\rm pl}$, they could then be safely ignored, and the mixing effects would be independent of the value of $m_a$ as indicated in Eq.~(\ref{eq_prob}$-$\ref{eq_EL}). Considering that the measured value of the time-integrated linear polarization ($\Pi_{L}$) we are focusing on is in the sub-MeV energy range and the plasma frequency of the intergalactic medium under our consideration is $\omega_{\rm pl} \simeq 1.17 \times 10^{-14}$ eV, as indicated by $E_{L}$, to ensure that photon-ALP mixing occurs within the strong mixing regime, we will use $g_{a\gamma} = 0.5 \times 10^{-11}$ GeV$^{-1}$ for $m_{a} \leq 10^{-14}$ eV as our benchmark model parameters.


\section{Results}
\label{Sec_results}

\begin{figure}[!tbp]
\begin{minipage}[t]{0.99\linewidth}
\centering
\includegraphics[width=1.0\textwidth]{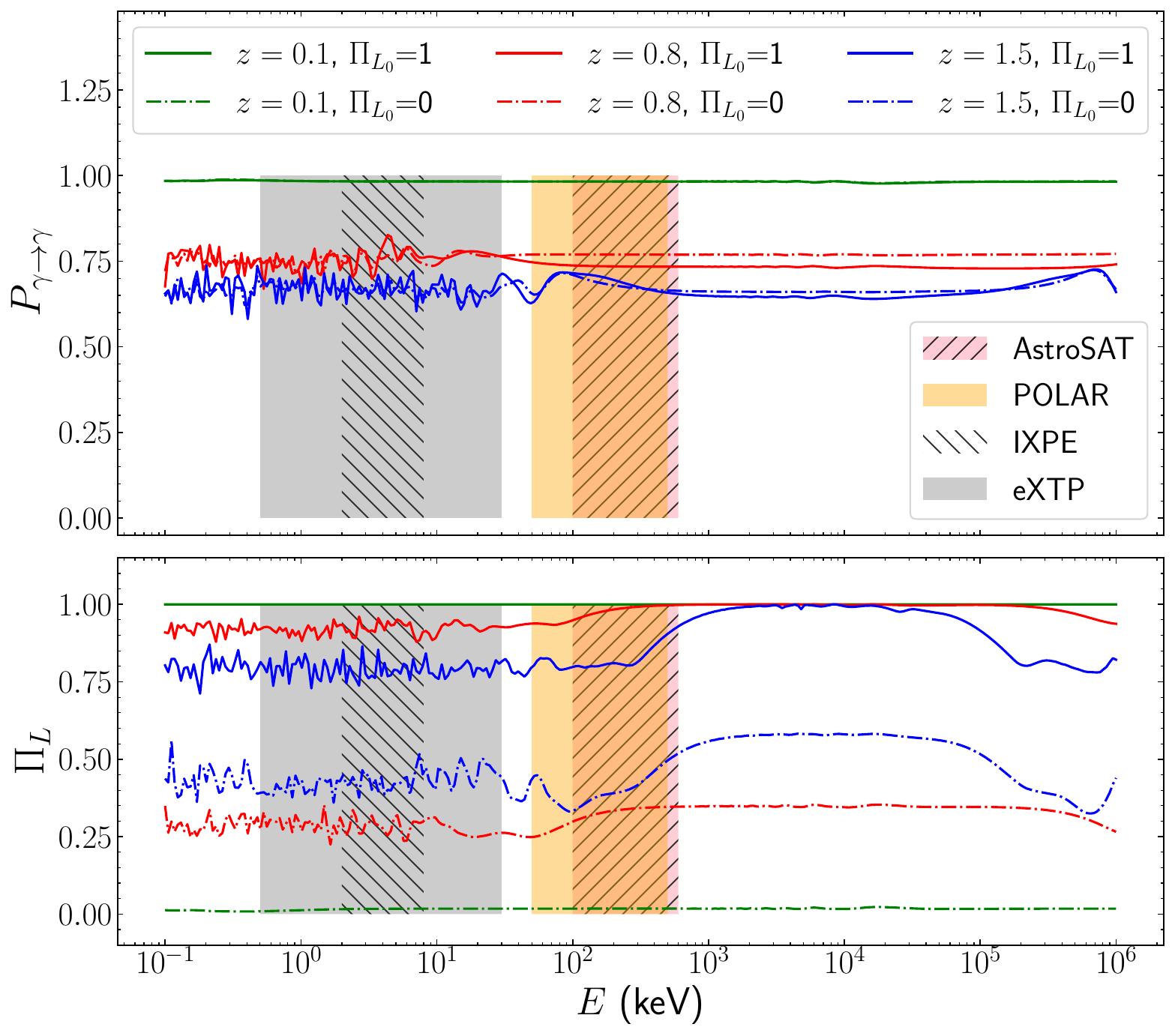}
\caption{Average final photon survival probability $P_{\gamma\rightarrow\gamma}$ (upper panel) and average final linear polarization degree $\Pi_{L}$ (lower panel) as a function of photon energy $E$ for ALP with $m_{a}$$~$=$~$$10^{-14}$$~$ eV and $g_{a\gamma}$$~$=$~0.5\times10^{-11}$$~$ GeV$^{-1}$. 
Green, red, and blue lines denote source redshifts of 0.1, 0.8, and 1.5, respectively. Solid and dash-dotted lines represent $\Pi_{L_{0}}$=1 and $\Pi_{L_{0}}$=0, respectively. Shadow zones denote energy bands for various polarimeters: AstroSAT (shaded pink), POLAR (orange), IXPE (shaded black), and eXTP (grey).}
\label{fig:pol_E_z-change}
\end{minipage}
\end{figure}


This section presents our results using the elaborated parameters set above. Firstly, we consider ALPs with $m_{a}=10^{-14}$ eV and $g_{a\gamma}=0.5\times10^{-11}$ GeV$^{-1}$ to explore the primary features of the potential effects induced by photon-ALP mixing in the extragalactic medium.  All calculations presented here have been conducted with mixing in ${\bf B}_{\rm IGM}$ and ${\bf B}_{\rm MW}$. However, we have examined sources from different lines of sight, e.g., Galactic latitudes $b$ ranging from 10$^{\circ}$, 40$^{\circ}$, to 60$^{\circ}$, and found no significant differences in the results. 
Therefore, in the sub-MeV energy band with the benchmark model we consider, variations induced by ${\bf B}_{\rm MW}$ can be disregarded. 

In Fig.$~$\ref{fig:pol_E_z-change}, we display the photon survival probability $P_{\gamma\rightarrow\gamma}(E)$ (upper panel) and the linear polarization degree $\Pi_{L}(E)$ (lower panel) for initially unpolarized photons ($\Pi_{L_{0}}=0$, dash-dotted lines) and fully linearly polarized photons ($\Pi_{L_{0}}=1$, solid lines) emitted at three typical redshifts, each indicated by a different color. $P_{\gamma\rightarrow\gamma}$ maintains its independence of photon energy E, as anticipated in cases of strong mixing. The photon flux remains unchanged for low redshift ($z$=0.1) while decreasing 20\% for medium redshift ($z$=0.8) and 40\% for high redshift ($z$=1.5) irrespective of their initial polarization states. The asymptotic trend of $P_{\gamma\rightarrow\gamma}$ approaching $2/3$ for high-redshift sources is foreseeable when photon-ALP beams traverse a sufficient number of domains \citep{GRBpol-IGM,GRBpol-IQUV}. 

$\Pi_{L}(E)$ exhibits similarities with the observed features in $P_{\gamma\rightarrow\gamma}(E)$ regarding energy. The modifications as defined by $\Delta_{\Pi_{L}} \equiv |\Pi_{L}-\Pi_{L_{0}}|$ are different for different initial polarization states assumed. For the low-redshift source (z=0.1), the modifications induced by the photon-ALP mixing are negligible as shown by the green lines in the bottom panel; the source maintains its initial polarization state. For sources at higher redshifts, photon-ALP mixing leads to a decrease in the polarization degree of initially fully linearly polarized photons, while it induces a certain degree of polarization to initially unpolarized photons. The magnitude of this alteration increases with higher redshift. In the keV to MeV energy range, at $z$=0.8, $\Delta_{\Pi_{L}}$ for $\Pi_{L_{0}}=1.0$ is about 0.1, while at $z$=1.5, it is approximately 0.2. At $z$=0.8, the polarization degree of unpolarized photons increases to 0.3, whereas at $z$=1.5, it increases to around 0.4. While these specific values correspond to the averaged values over 50 realizations of the orientations of the series domains of $\mathbf{B}_{\text{IGM}}$, these case studies have revealed the trend that alterations to $\Pi_{L_{0}}$ could be significant for our benchmark model. We also note that for photons with energies greater than MeV, these changes in the unpolarized case become more pronounced. As our study focuses on the sub-MeV energy range, we do not delve further into this aspect. 

\begin{figure}[!tbp]
\begin{minipage}[t]{0.99\linewidth}
\centering
\includegraphics[width=1.0\textwidth]{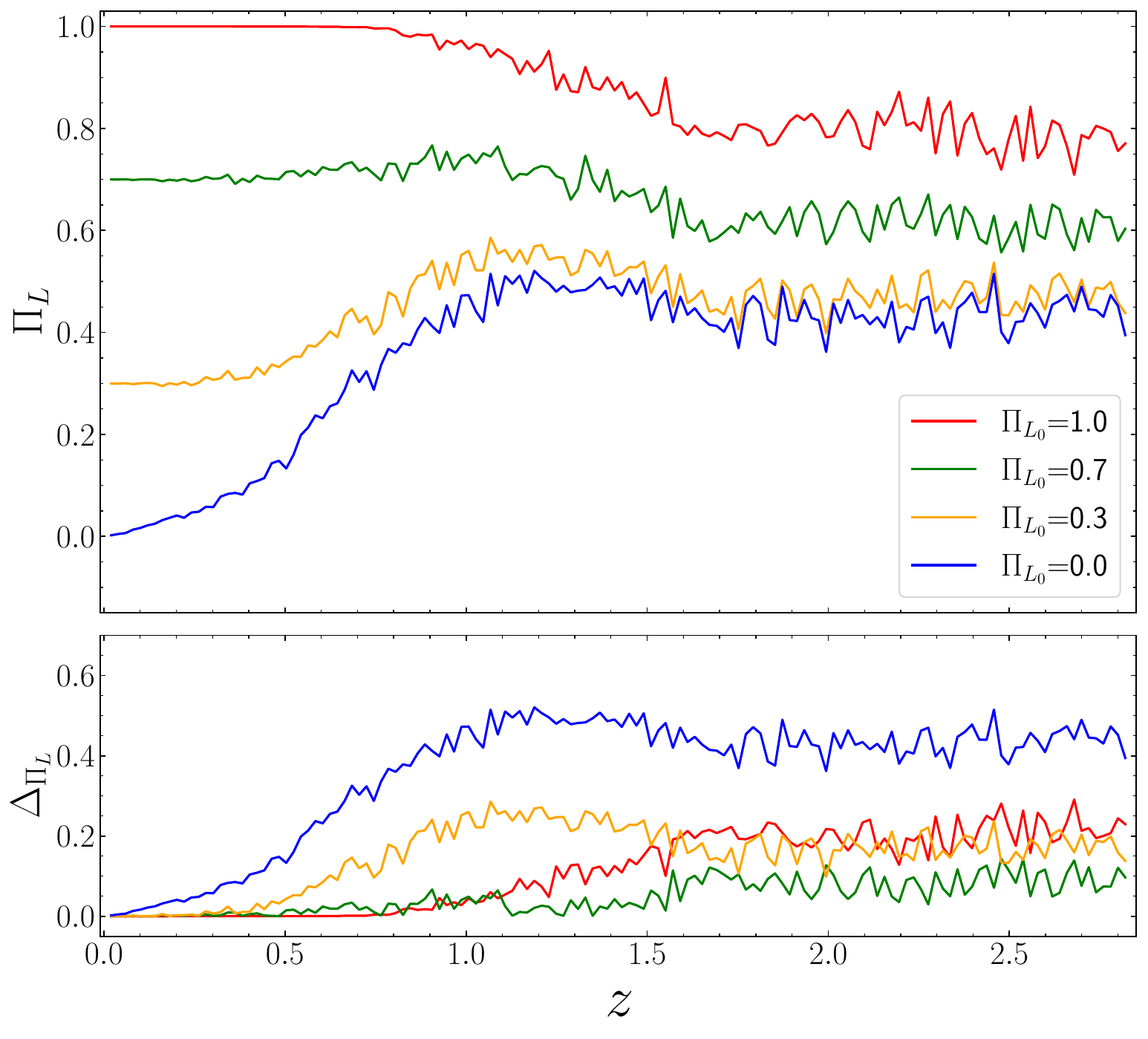}
\caption{Average final $\Pi_{L}$ (upper panel) and $\Delta_{\Pi_{L}}$ (lower panel) as a function of the source redshift $z$ for ALP with  $m_{a}$=$10^{-14}$ eV and $g_{a\gamma}$=$0.5\times10^{-11}$ GeV$^{-1}$ and photon energy $E=350$ keV. Colors indicate different $\Pi_{L_{0}}$: 1.0 (red), 0.7 (green), 0.3 (orange), 0.0 (blue).}
\label{fig:pol_z_pin-change}
\end{minipage}
\end{figure}

\begin{figure*}[!tbp]
\centering
\includegraphics[width=1.0\textwidth]{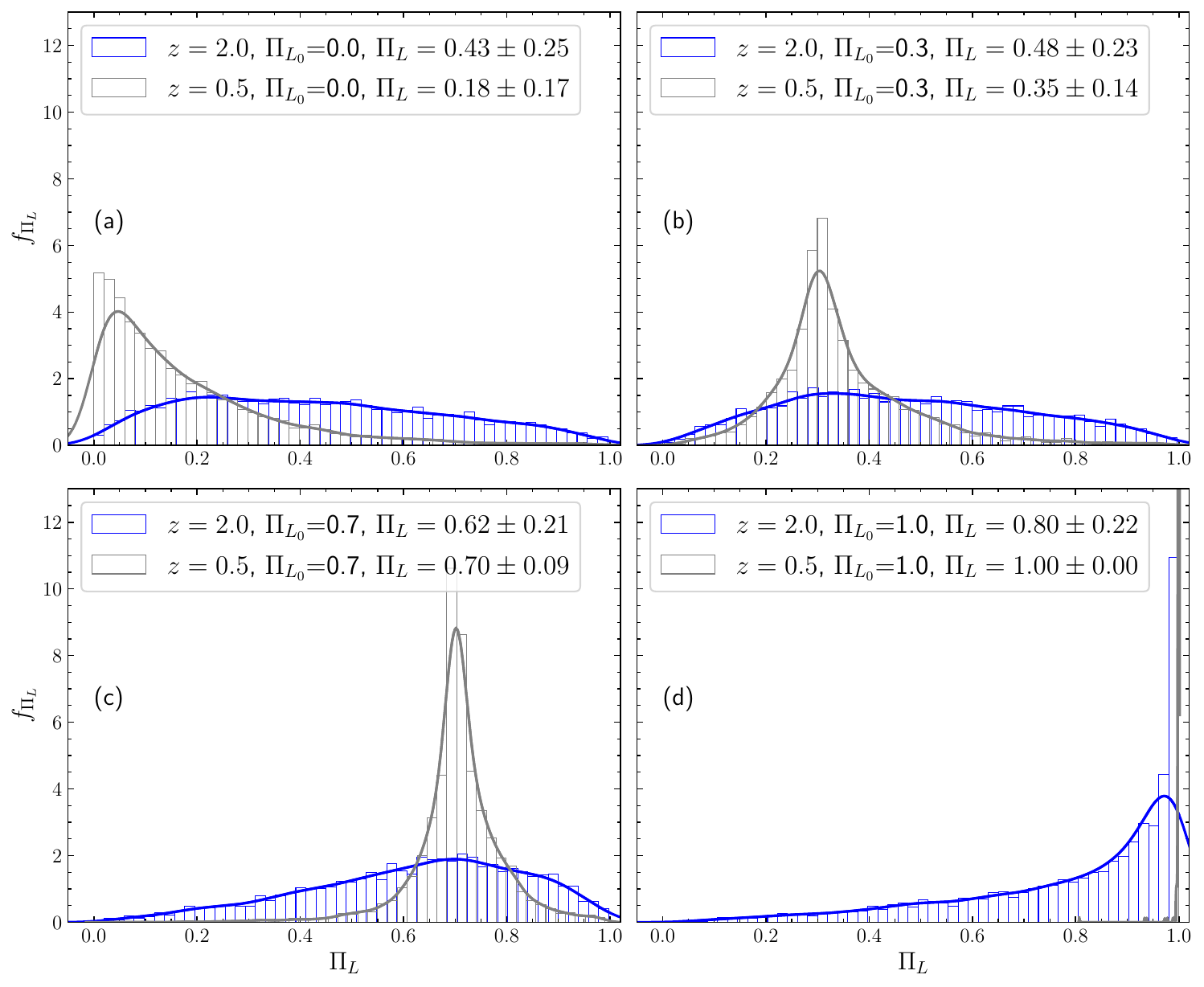}
\caption{Probability density function $f_{\Pi_L}$ of $\Pi_L$ for ALP with $m_{a}$=$10^{-14}$ eV and $g_{a\gamma}$=$0.5\times10^{-11}$ GeV$^{-1}$ and photon energy $E=350$ keV. Blue and grey represent sources at redshift $z=2.0$ and 0.5, respectively. In (a), (b), (c), and (d), for $z=2.0$, the mean final $\Pi_{L}$=$0.43\pm0.25$ ($\Pi_{L_{0}}=0.0$), $\Pi_{L}$=$0.48\pm0.23$ ($\Pi_{L_{0}}=0.3$), $\Pi_{L}$=$0.62\pm0.21$ ($\Pi_{L_{0}}=0.7$), $\Pi_{L}$=$0.80\pm0.22$ ($\Pi_{L_{0}}=1.0$); for $z=0.5$, $\Pi_{L}$=$0.18\pm0.17$ ($\Pi_{L_{0}}=0.0$), $\Pi_{L}$=$0.35\pm0.14$ ($\Pi_{L_{0}}=0.3$), $\Pi_{L}$=$0.70\pm0.09$ ($\Pi_{L_{0}}=0.7$), $\Pi_{L}$=$1.00\pm0.00$ ($\Pi_{L_{0}}=1.0$).}
\label{fig.pol_PDF}
\end{figure*}

In Fig.$~$\ref{fig:pol_E_z-change}, we have also marked the relevant energy bands of the high-energy polarimeters as shadowed areas. The shaded pink, orange, shaded black, and grey shadow zone represent AstroSAT, POLAR, IXPE, and eXTP, respectively. As illustrated in Table.~\ref{tab:Measurements}, most of the measured $\Pi_{L}$ with possible redshift identified are obtained by AstroSAT, which covers the 100-600 keV energy band. Hence, we will choose the middle-energy $E=350$ keV to account for the measured photon energy and we further explore the alternations to $\Pi_{L_{0}}$ induced by photon-ALP mixing for this specific energy.

As analytically deduced in Ref.$~$\citep{GRBpol-IQUV}, there is an asymptotic behavior for $\Pi_{L}$ when photons emitted from high-redshift GRBs are initially fully linearly polarized. Here, we present our numerical results of $\Pi_{L}(z)$ and $\Delta_{\Pi_{L}}(z)$ for a range of  $\Pi_{L_{0}}$ values as indicated by different colors in Fig.$~$\ref{fig:pol_z_pin-change}. In our results, this asymptotic behavior occurs across all the assumed $\Pi_{L_{0}}$ values. For $\Pi_{L_{0}}=0.0$ and $\Pi_{L_{0}}=0.3$, representing low initial polarization degrees, $\Delta_{\Pi_{L}}$ exhibits a significant increase starting at a redshift of approximately 0.5, reaching a maximum deviation close to 0.5, and stabilizing after redshift 1.7. For $\Pi_{L_{0}}=0.7$ and $\Pi_{L_{0}}=1.0$, representing high initial polarization degrees, visible increments begin around a redshift of approximately 1.0, with a maximum deviation close to 0.2, also stabilizing after redshift 1.7. From the upper panel, we can see that for sources at redshifts greater than 0.5, nearly all the photons exhibit $\Pi_{L}$ above 0.2, regardless of their initial polarization state. However, as indicated in Fig.$~$\ref{fig:pol_hist}, 12 out of 50 observed GRBs have a measured $\Pi_{L}$ less than 0.2, with none having a confirmed redshift yet. Further confirmation of the distances of these sources could potentially impose strong constraints on the photon-ALP mixing.

The averaged values of $\Pi_{L}$ presented above are calculated across 50 realizations of the random magnetic field configurations of $\mathbf{B}_{\text{IGM}}$. 
It is important to note that each photon-ALP beam can only encounter one configuration at a time, and individual realizations do not hold representative significance. However, through computing numerous realizations, we can derive its statistical characteristics. Therefore, we analyze the probability density function $f_{\Pi}$ associated with $\Pi_{L}$ based on 5000 realizations of random magnetic field configurations. The resulting $f_{\Pi}$ and the corresponding mean value and standard deviation of $\Pi_{L}$ are presented in Fig.$~$\ref{fig.pol_PDF}, where (a), (b), (c), and (d) represent $\Pi_{L_{0}}~=~$0.0, 0.3, 0.7, and 1.0, respectively. For redshift $z=0.5$ (grey color), all $f_{\Pi}$ presents a peak approaching its initial polarization state. For low initial polarization degrees, the mean values of $\Pi_{L}$ show deviations as labeled in the figure, $\Pi_{L}$=$0.18\pm0.17$ for $\Pi_{L_{0}}=0.0$ and $\Pi_{L}$=$0.35\pm0.14$ for $\Pi_{L_{0}}=0.3$. For redshift $z=2.0$ (blue color), all $f_{\Pi}$ distributions appear rather flat, showing no clustering with the initial polarization state. The mean and 1$\sigma$ values of $\Pi_{L}$ are as follows: $0.43\pm0.25$ for $\Pi_{L_{0}}=0.0$, $0.48\pm0.23$ for $\Pi_{L_{0}}=0.3$, $0.62\pm0.21$ for $\Pi_{L_{0}}=0.7$, and $0.80\pm0.22$ for $\Pi_{L_{0}}=1.0$. Thus, the photon-ALP mixing in our benchmark model may smear out the initial polarization state of photons emitted from high-redshift sources. These results are consistent with the asymptotic behavior discussed in Fig.$~$\ref{fig:pol_z_pin-change}, indicating that averaging over 50 realizations is sufficient to capture the asymptotic behavior.


\begin{figure}[!tbp]
\begin{minipage}[t]{0.99\linewidth}
\centering
\includegraphics[width=1.0\textwidth]{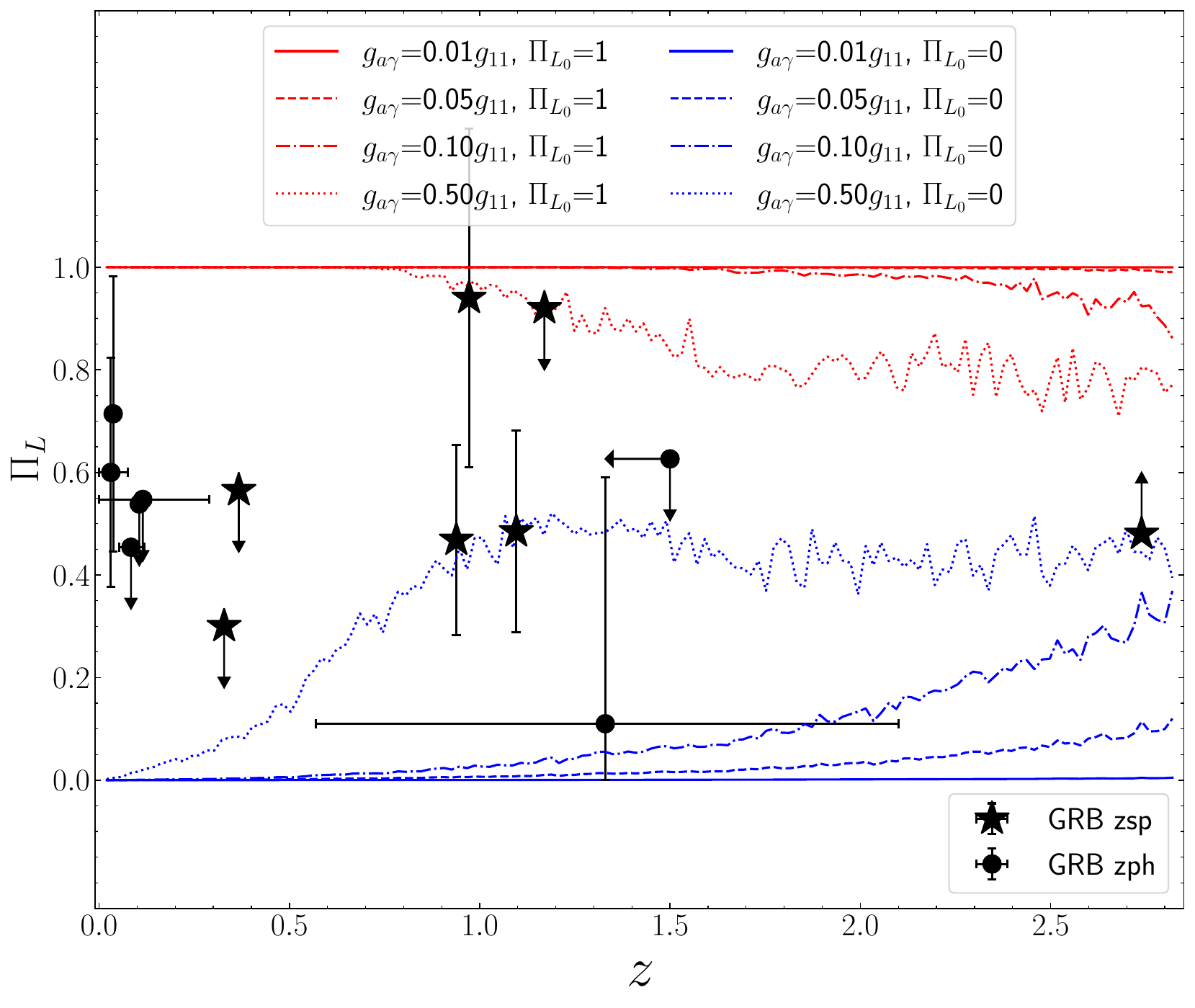}
\caption{Average final $\Pi_{L}$ as a function of the source redshift $z$ for ALP with $m_{a}$=$10^{-14}$ eV and photon energy $E=350$ keV. Red/blue line represents $\Pi_{L_{0}}$=1.0/0.0. The different line styles represent different $g_{a\gamma}$, i.e., $0.01\times10^{-11}$ GeV$^{-1}$ (solid line), $0.05\times10^{-11}$ GeV$^{-1}$ (dashed line), $0.1\times10^{-11}$ GeV$^{-1}$ (dash-dotted line), $0.5\times10^{-11}$ GeV$^{-1}$ (dotted line), and $g_{11}=10^{-11}$ GeV$^{-1}$. The data points are the sources listed in Table \ref{tab:Measurements}. The pentagon/dot denotes sources with/without spectroscopic redshift.}
\label{fig:pol_z_m13}
\end{minipage}
\end{figure}

\begin{figure*}[!tbp]
\centering
\includegraphics[width=1.0\textwidth]{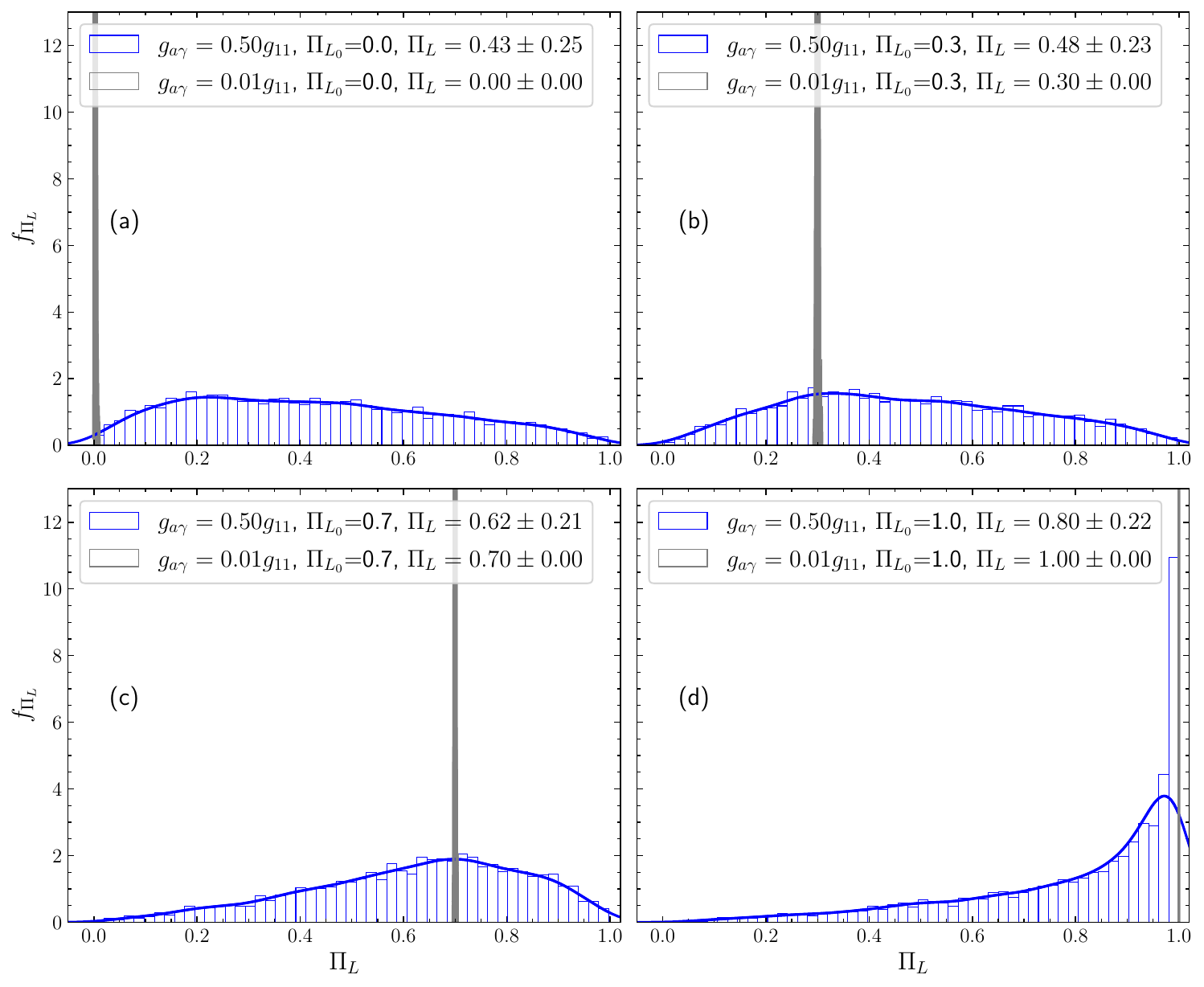}
\caption{Probability density function $f_{\Pi_L}$ of $\Pi_L$ for ALPs with $m_{a}$=$10^{-14}$ eV and photon energy $E=350$ keV assuming a source at $z=2$. $g_{11}=10^{-11}$ GeV$^{-1}$. Blue and grey color represent $g_{a\gamma}$=$0.5\times10^{-11}$ GeV$^{-1}$ and $0.01\times10^{-11}$ GeV$^{-1}$, respectively. 
}
\label{fig.pol_PDF_g13}
\end{figure*}

\begin{figure}[!tbp]
\begin{minipage}[t]{0.99\linewidth}
\centering
\includegraphics[width=1.0\textwidth]{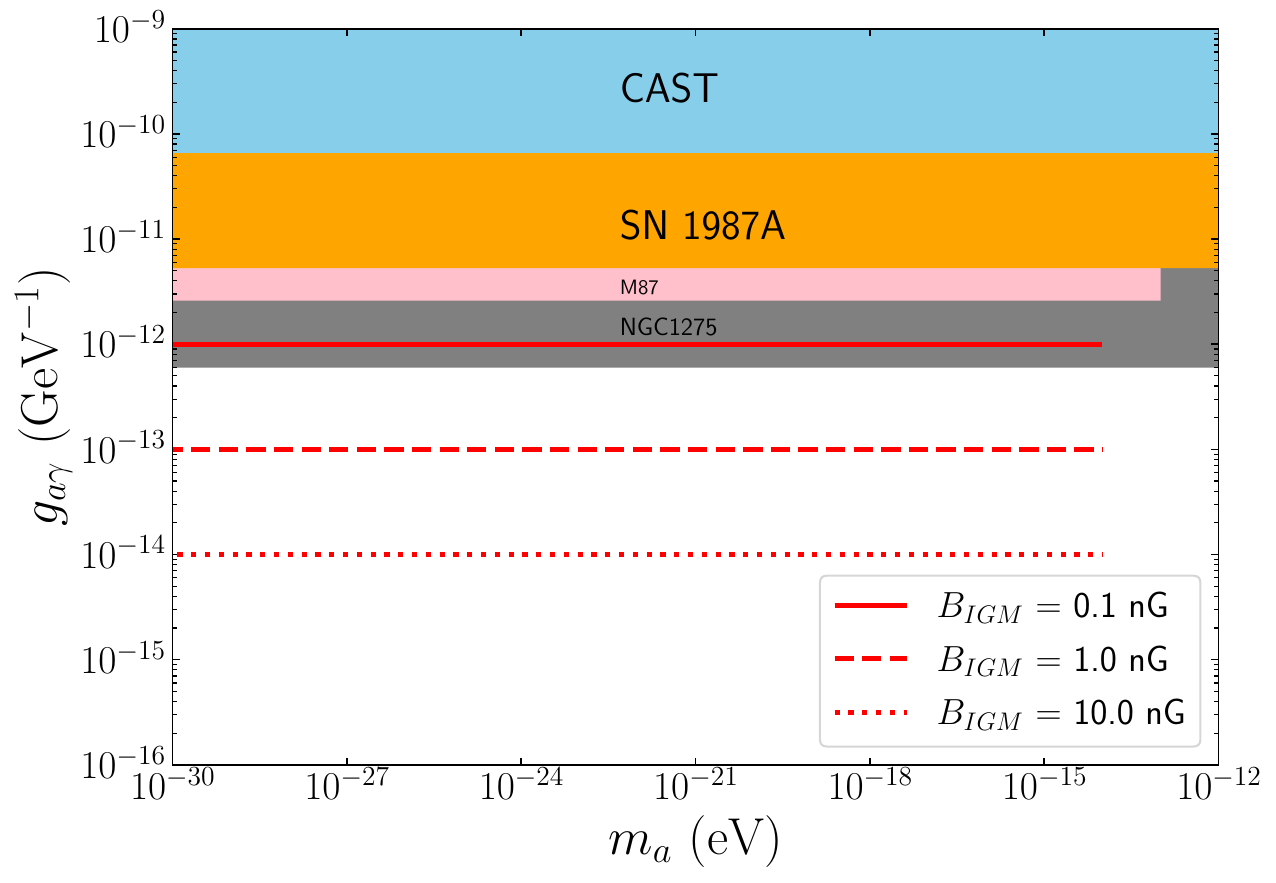}
\caption{Bounds on $g_{a\gamma}$ as $\Delta_{\Pi_L}$ approaches to zero. The solid/dashed/dotted line corresponds to $\bf B$$_{\rm IGM}$$~$=$~$0.1/1.0/10.0$~$nG. The colored regions represent previously obtained bounds.}
\label{fig:ALP_limit}
\end{minipage}
\end{figure}

Our above calculations used a benchmark model of ALPs with $m_{a}=10^{-14}$ eV and $g_{a\gamma}=0.5\times10^{-11}$ GeV$^{-1}$. Since $m_{a}=10^{-14}$ eV is smaller than $\omega_{\rm pl} \simeq 1.17 \times 10^{-14}$ eV, the ALP mass term could be safely dropped from the mixing term. The photon-ALP mixing effect would be independent of the value of ALP mass for $m_{a}\lesssim10^{-14}$ eV. For such kind of low-mass ALPs, we further
explore to what extent, the alternations to $\Pi_{L_{0}}$ are small enough to be negligible given different $g_{a\gamma}$ values. Fig.$~$\ref{fig:pol_z_m13} shows the average final linear polarization degree $\Pi_{L}$ as a function of source redshift for two scenarios: initially unpolarized photons ($\Pi_{L_{0}}=0$, blue lines) and fully polarized photons ($\Pi_{L_{0}}=1$, red lines). 
Different line styles represent different values of $g_{a\gamma}$. As the value of $g_{a\gamma}$ goes from $0.5\times10^{-11}$ GeV$^{-1}$ (dotted lines), $0.1\times10^{-11}$ GeV$^{-1}$ (dash-dotted lines), $0.05\times10^{-11}$ GeV$^{-1}$ (dashed lines) down to $0.01\times10^{-11}$ GeV$^{-1}$ (solid lines), the modifications induced by mixing tend to decrease as expected. For the smallest value of $g_{a\gamma}$ considered in this study, sources at any redshift may retain their initial linear polarization state, irrespective of the assumed values of $\Pi_{L_{0}}$. This feature can be further confirmed in Fig.$~$\ref{fig.pol_PDF_g13}, where we present $f_{\Pi}$ (based on 5000 realizations of random magnetic field configurations) along with the corresponding mean value and standard deviation of $\Pi_{L}$. In comparison to our benchmark model, $f_{\Pi}$ for $g_{a\gamma}=0.01\times10^{-11}$ GeV$^{-1}$ peaks at the exact value of the initial polarization state, regardless of $\Pi_{L_{0}}$, even when the redshift is as high as 2.0. The effect induced by photon-ALP mixing could be safely ignored. This result, in turn, leads us to conclusion that the strongest constraints on the upper limit of $g_{a\gamma}$ reach the order of $10^{-13}$ GeV$^{-1}$.

The sources listed in Table \ref{tab:Measurements} are marked as
the data points in Fig.$~$\ref{fig:pol_z_m13}, where
the pentagon (dot) denotes sources with spectroscopic redshift (possible photometric redshifts or upper limits). We have examined the correlation between the measured $\Pi_{L}$ and the redshifts of GRBs as shown in Fig.$~$\ref{fig:pol_Zhost}, indicating the lack of statistical behavior probably due to the limited number of the data points and the larger uncertainties either in measured $\Pi_{L}$ or in the redshift determination. Constraints may be set from individual cases. A plausibly extremely useful one could be the lowest $\Pi_{L}$ (GRB061122A) in Fig.$~$\ref{fig:pol_z_m13}, but unfortunately, the redshift is a photometric one resembling its possible host candidate galaxy, which severely suffers from larger uncertainties that require further confirmation. A more precise measurement of the redshift with a low value of $\Pi_{L}$ could impose strong constraints on the photon-ALP mixing term $\Delta_{a\gamma} \equiv \frac{1}{2} g_{a\gamma} B_{\rm IGM}$. For a fixed value of $\bf B$$_{\rm IGM}$$~$=$~$1$~$nG, GRB061122A-like polarization measurements at least lead to the exclusion of our benchmark model $g_{a\gamma}=0.5\times10^{-11}$ GeV$^{-1}$ for $m_{a}\lesssim10^{-14}$ eV. 

To ensure that the photon maintains its initial polarization state after propagation in the extragalactic space, the strongest constraints on the upper limit of $g_{a\gamma}$ possibly reach the order of $10^{-13}$ GeV$^{-1}$ for $\bf B$$_{\rm IGM}$$~$=$~$1$~$nG, which means a requirement of the mixing term $\Delta_{a\gamma}=1.5\times 10^{-2}(g_{a\gamma}/10^{-11}\rm GeV^{-1})(B_{\rm IGM}/\rm nG)\ \rm Mpc^{-1} \leq 1.5\times 10^{-4}\ \rm Mpc^{-1}$. Fig.$~$\ref{fig:ALP_limit} illustrates how the constraint on $g_{a\gamma}$ changes when ${\bf B}_{\rm IGM}$ is varied to different values. The solid/dashed/dotted line corresponds to $\bf B$$_{\rm IGM}$$~$=$~$0.1/1.0/10.0$~$nG. The colored regions represent several important constraints \citep{SN1987_limit2015, CAST_limit, Marsh_2017, Reynolds_2020}. In the future, accurate measurements of ${\bf B}_{\rm IGM}$ could more precisely limit $g_{a\gamma}$, or conversely, measurements of $g_{a\gamma}$ could in turn provide independent checks on ${\bf B}_{\rm IGM}$.

\section{Conclusion and Discussion}
\label{Sec_conclusion}
In this work, we first examine the statistical characteristics of the measured time-integrated $\Pi_{L}$ from different polarimetric missions focusing on sub-MeV emissions. Our analysis reveals a diverse distribution of $\Pi_{L}$, which currently shows no correlation with spectral parameters or properties of candidate host galaxies. A thorough investigation of the redshift recognition has also been conducted to achieve a sample of 14 GRBs with both measured $\Pi_{L}$ and redshift. The corresponding Spearman correlation coefficient is $r_{s}$~$=$~$-0.18\pm0.45$, indicating a lack of statistical behavior within the current sample due to large uncertainties in measured  $\Pi_{L}$ and redshift recognitions. Then we delve into the possible alternations $\Delta_{\Pi_{L}} \equiv |\Pi_{L}-\Pi_{L_{0}}|$ and present the primary features of the potential effects induced by photon-ALP mixing based on our fiducial model which satisfy current constraints on the ALP parameter space and ${\bf B}_{\rm IGM} $. Photon-ALP mixing in the extragalactic space would further exaggerate the diversity of $\Pi_{L}$ and smear out the initial polarization state of the photons emitted by GRBs at high redshifts. To maintain the initial polarization state of a photon after traversing extragalactic space, the most stringent constraints on the upper limit of $g_{a\gamma}$ are on the order of $10^{-13}$ GeV$^{-1}$ for $\mathbf{B}_{\text{IGM}} = 1$ nG, corresponding to a requirement that the mixing term $\Delta{a\gamma}$ be no greater than $1.5\times 10^{-4}\ \text{Mpc}^{-1}$. In the future, better measurements of $\mathbf{B}_{\text{IGM}}$ will impose secure limits on $g_{a\gamma}$, and conversely, good determination of $g_{a\gamma}$ will  provide independent constraints on $\mathbf{B}_{\text{IGM}}$.

The lowest measured value ($\sim$11\%) of $\Pi_{L}$ with redshift recognization is GRB061122A but its photometric redshift ($z\sim 1.3$) suffers larger uncertainties that require further confirmation. Meanwhile, it is notable that 12 out of 50 observed GRBs have a measured $\Pi_{L}$ less than 0.2, with none of them having a confirmed redshift yet. Further confirmation of the distances of these low-$\Pi_{L}$ sources could potentially impose strong constraints on the photon-ALP mixing in extragalactic space. The ongoing polarimeter (CZTI) on-board AstroSAT is expected to increase the number of GRBs with measured $\Pi_{L}$ and improve localization accuracy \citep{ Chattopadhyay2022}.
The forthcoming POLAR-2 \citep{POLARII} and the Large Area Burst Polarimeter (LEAP; \citep{LEAP_2021}), operating in the energy range of $10-1000$ keV, will significantly increase the number of GRBs observed across various redshifts. The statistical enhancement in the detection of high-energy polarization from GRBs, along with more accurate redshift verifications, is anticipated and thus better constraints on the mixing term will be achieved in the near future.

The most precise measurement of $\Pi_{L}$ of GRBs is derived from observations of GRB221009A by IXPE. The small value of $\Pi_{L}=0.06\pm0.03$ at 2--8\,keV is attributed to dust-induced polarization within the Milky Way (indicated by days time-lag compared to its prompt phase), rendering it unsuitable for constraining mixing in IGM \citep{2023ApJ...946L..21N}. Persistent observations conducted by IXPE on blazars with peak frequencies in the soft X-ray band or AGNs exhibiting thermal emission from their surrounding coronae at high redshifts would impose limits on the mixing term through a similar methodology, although its newly planned observations continue to prioritize sources at low redshifts. We note that planned missions such as the Enhanced X-ray Timing and Polarimetry mission (eXTP; \citep{eXTP_2016}) and the X-ray Polarization Probe (XPP; \citep{XPP_2019}) with higher sensitivity will extend detection towards high-redshift sources within energy ranges of $0.5-30$ keV and $0.2-60$ keV, respectively. In addition, future polarimeters, such as the Compton Spectrometer and Imager (COSI; \citep{COSI_2019}) and the All-Sky Medium Energy Gamma-ray Observatory (AMEGO; \citep{AMEGO_2019}), targeting higher energy ranges ($100$ keV$-5$ MeV), are prepared to investigate the MeV polarization features, as shown in the lower panel of Fig.$~$\ref{fig:pol_E_z-change}. These ongoing and planned missions will certainly provide a much larger sample for the investigation in this work and hopefully clear results would be obtained.

\section*{Acknowledgements}

This work is supported by the National Natural Science Foundation of China under Nos. 11890692, 12133008, 12221003. We acknowledge the science research grant from the China Manned Space Project with No. CMS-CSST-2021-A04. We have utilized the following softwares: NumPy \citep{numpy2020}, Astropy \citep{astropy:2013, astropy:2018, astropy:2022}, Matplotlib \citep{matplotlib2007}, gammaALPs \citep{gammaALPs2022}, CIGALE \citep{cigale2019, cigale2020, cigale2022}, LePhare \citep{lephare1999, lephare2006}.


\end{document}